\documentclass[aps,pra,twocolumn,notitlepage,10pt,superscriptaddress,longbibliography,nofootinbib]{revtex4-2}

\usepackage{graphicx}
\usepackage{times,bbm,amsmath,amssymb}

\usepackage{hyperref}
\usepackage{xcolor}
\usepackage{cleveref}
\crefname{figure}{Fig.}{Figs.}
\Crefname{figure}{Figure}{Figures}

\usepackage{graphicx}
\usepackage{dcolumn}%
\usepackage{bm}
\usepackage{soul}

\usepackage{nicefrac, xfrac}
\usepackage{pifont}
\usepackage{bbold}
\usepackage{physics}

\usepackage{multirow,tabularx,booktabs}
\usepackage{makecell}

\usepackage{easyReview}

\usepackage{parskip}
\usepackage{libertine}[newtxmath]

\usepackage{xr}
\newcommand{\calO}{\mathcal{O}}

\newcommand{\calW}{\mathcal{W}}

\newcommand{\bs}[1]{\boldsymbol{#1}}
\newcommand{\on}[1]{\operatorname{#1}}
\newcommand{\parTitle}[1]{\noindent\emph{#1} ---}

\newcommand{\E}[1]{$\mathsf{E#1}$}

\newcommand{\bstildemu}{\bs{\tilde{\mu}}}
\newcommand{\sfE}[1]{$\boldsymbol{\mathsf{E#1}}$}

\newcommand{\sfR}[1]{$\mathsf{R#1}$}
\usepackage{xspace}
\newcommand{\resOptimal}{$\mathsf{R1}$\xspace}
\newcommand{\resFakeOptimal}{$\mathsf{R2}$\xspace}
\newcommand{\resRandom}{$\mathsf{R3}$\xspace}

\definecolor{azure}{rgb}{0.0, 0.3, 1.0}

\setcitestyle{numbers,round}

\begin{document}

\title{Quantum reservoir computing for photonic entanglement witnessing}

\let\comma,

\author{Danilo Zia}\thanks{These authors contributed equally to this work}
\affiliation{Dipartimento di Fisica - Sapienza Università di Roma\comma{} P.le Aldo Moro 5\comma{} I-00185 Roma\comma{} Italy}
\author{Luca Innocenti}\thanks{These authors contributed equally to this work}
\affiliation{Universit\`a degli Studi di Palermo\comma{} Dipartimento di Fisica e Chimica - Emilio Segr\`e\comma{} via Archirafi 36\comma{} I-90123 Palermo\comma{} Italy}
\author{Giorgio Minati}
\affiliation{Dipartimento di Fisica - Sapienza Università di Roma\comma{} P.le Aldo Moro 5\comma{} I-00185 Roma\comma{} Italy}
\author{Salvatore Lorenzo}
\affiliation{Universit\`a degli Studi di Palermo\comma{} Dipartimento di Fisica e Chimica - Emilio Segr\`e\comma{} via Archirafi 36\comma{} I-90123 Palermo\comma{} Italy}
\author{Alessia Suprano}
\affiliation{Dipartimento di Fisica - Sapienza Università di Roma\comma{} P.le Aldo Moro 5\comma{} I-00185 Roma\comma{} Italy}
\author{Rosario Di Bartolo}
\affiliation{Dipartimento di Fisica - Sapienza Università di Roma\comma{} P.le Aldo Moro 5\comma{} I-00185 Roma\comma{} Italy}
\author{Nicol\`{o} Spagnolo}
\affiliation{Dipartimento di Fisica - Sapienza Università di Roma\comma{} P.le Aldo Moro 5\comma{} I-00185 Roma\comma{} Italy}
\author{Taira Giordani}
\affiliation{Dipartimento di Fisica - Sapienza Università di Roma\comma{} P.le Aldo Moro 5\comma{} I-00185 Roma\comma{} Italy}
\author{Valeria Cimini}
\affiliation{Dipartimento di Fisica - Sapienza Università di Roma\comma{} P.le Aldo Moro 5\comma{} I-00185 Roma\comma{} Italy}
\author{G. Massimo Palma}
\affiliation{Universit\`a degli Studi di Palermo\comma{} Dipartimento di Fisica e Chimica - Emilio Segr\`e\comma{} via Archirafi 36\comma{} I-90123 Palermo\comma{} Italy}
\author{Alessandro Ferraro}
\affiliation{Quantum Technology Lab\comma{} Dipartimento di Fisica Aldo Pontremoli\comma{} Universit\`a degli Studi di Milano\comma{} I-20133 Milano\comma{} Italy}
\author{Fabio Sciarrino }\email{fabio.sciarrino@uniroma1.it}
\affiliation{Dipartimento di Fisica - Sapienza Università di Roma\comma{} P.le Aldo Moro 5\comma{} I-00185 Roma\comma{} Italy}
\author{Mauro Paternostro}\email{mauro.paternostro@unipa.it}
\affiliation{Universit\`a degli Studi di Palermo\comma{} Dipartimento di Fisica e Chimica - Emilio Segr\`e\comma{} via Archirafi 36\comma{} I-90123 Palermo\comma{} Italy}
\affiliation{Centre for Quantum Materials and Technologies\comma{} School of Mathematics and Physics\comma{} Queen's University Belfast\comma{} BT7 1NN\comma{} United Kingdom}

\begin{abstract}
Accurately estimating properties of quantum states, such as entanglement, while essential for the development of quantum technologies, remains a challenging task. Standard approaches to property estimation rely on detailed modeling of the measurement apparatus and a priori assumptions on their working principles. Even small deviations can greatly affect reconstruction accuracy and prediction reliability. Here, we demonstrate that quantum reservoir computing embodies a powerful alternative for witnessing quantum entanglement and, more generally, estimating quantum features from experimental data. We leverage the orbital angular momentum of photon pairs as an ancillary degree of freedom to enable informationally complete single-setting measurements of their polarization. Our approach does not require fine-tuning or refined knowledge of the setup, at the same time outperforming conventional approaches. It automatically adapts to noise and imperfections while avoiding overfitting, ensuring robust reconstruction of entanglement witnesses and paving the way to the assessment of quantum features of experimental multiparty states.

\end{abstract}

\maketitle

\section{Introduction}

The accurate estimation of quantum states properties remains a pivotal achievement for the advancement of quantum technologies. 
The golden standard in this context is embodied by quantum state tomography, whose goal is fully reconstructing a given quantum state~\cite{dariano2003quantum,paris2004quantum}, as well as quantum state estimation and certification protocols~\cite{kliesch2021theory,eisert2020quantum,gebhart2023learning,yu2022statistical,zambrano2024certification}.
More recently, \textit{classical shadows} have shown how to efficiently estimate several properties of unknown quantum states, even for large state dimensions, provided suitable measurement strategies are employed~\cite{huang2020predicting,innocenti2023shadow,acharya2021informationally,nguyen2022optimising,elben2023randomized}.

From an experimental perspective, we recognize two broad classes of measurement strategies.
On the one hand, \emph{single-setting} schemes~\cite{bian2015realization,stricker2022experimental,an2024efficient,oren2017quantum,smania2020experimental,fischer2022ancilla}
employ a fixed experimental apparatus, thereby avoiding device reconfigurations.
On the other hand, protocols based on multiple measurement bases --- including standard tomographic approaches as well as strategies based on random measurements and shadow tomography~\cite{magesan2012characterizing,proctor2019direct,struchalin2021experimental,zhang2021experimental,lib2025experimental,antesberger2024efficient} 
--- can sometimes reduce resource requirements, but necessitate frequent apparatus adjustments, intensifying the challenge of device calibration.
Despite these operational differences, all such strategies hinge critically on an accurate modeling of both the dynamical evolution and the measurement stage: unaccounted-for features of the apparatus will inevitably lead to systematic errors and unreliable outcomes.

Several machine learning (ML) approaches have also been proposed to enhance the performance of state estimation protocols~\cite{lu2018separability,palmieri2020experimental,harney2020entanglement,lohani2020machine,gebhart2020neural,giordani2020machine,suprano2021dynamical,suprano2021enhanced,schmale2022efficient,greenwood2023machine,cimini2023deep,zia2023regression,krisnanda2024experimental,ahmed2021quantum,innan2024quantum}.
ML-based methods, however, are often deficient in interpretability, and the opacity of the underlying models can complicate assessing the reliability of their results.
These issues are further exacerbated when the goal is witnessing nonclassical features of quantum states, such as  entanglement~\cite{pezze2016witnessing,dirkse2020witnessing,rosset2012imperfect}.

Here, we report the experimental realization of an estimation platform based on a memoryless variant of quantum reservoir computing, known as quantum extreme learning machines (QELMs)~\cite{mujal2021opportunities,huanggh2011extreme,krisnanda2023tomographic,assil2025entanglement, innocenti2023potential,ghosh2019quantum}, intending to reconstruct entanglement witnesses on two-qubit entangled states encoded in the polarization degrees of freedom of photon pairs. More specifically, we employ a double quantum-walk-based apparatus in the polarization and orbital angular momentum (OAM) degrees of freedom~\cite{allen1992orbital,yao2011orbital} of two-photon input states. The quantum walks are implemented using passive and active optical elements, specifically polarization waveplates and q-plates (QPs)~\cite{marrucci2006optical}.
This architecture allows embedding the polarization information into the much larger OAM space, whose measurement then yields an informationally-complete measurement of the input polarization states~\cite{suprano2024experimental}. 
The choice of this architecture is motivated by its ability to embed information into the large OAM space using simple control devices such as polarization waveplates and q-plates. This apparatus is not expected to be ideal for state reconstruction, since its effective measurement differs substantially from the symmetric measurements --- such as MUBs and SIC-POVMs --- known to be optimal for such tasks~\cite{huang2020predicting,innocenti2023shadow}. Moreover, accurately modeling all components of the apparatus and its noise sources is challenging. Nevertheless, as we will show, the proposed QELM strategy enables effective reconstruction even under such non-ideal conditions. This provides a strong proof-of-principle demonstration that our strategy can enhance reconstruction performances across a broad range of experimental scenarios, extending far beyond the context of photonics.

Despite its memoryless nature, the resulting QELM protocol overcomes several limitations of existing approaches by offering key advantages. In particular: 
{\bf (1)} it eliminates the need for prior modeling of the measurement apparatus, thereby completely circumventing the challenges associated with fine-tuning and calibration;
{\bf (2)} its straightforward training procedure --- amenable to rigorous theoretical analysis --- ensures reliable and robust performances;
{\bf (3)} by operating within a linear supervised machine learning framework, it sidesteps the interpretability and overfitting issues often encountered with more complex architectures.
Our QELM-based approach naturally aligns with the paradigm of self-calibrating strategies~\cite{mogilevtsev2012self,keith2018joint,zhang2020experimental} by dynamically learning the optimal way to post-process measurement data on-the-fly, thus mitigating the risk of miscalibration without extensive pre-calibration. Overall, the simplicity and generality of our approach make it adaptable to a wide range of experimental platforms and measurement devices, and advantageous in any situation where characterizing the state-preparation part of the experimental setup is easier than accurately modeling the whole evolution and measurement, including all sources of noise and misalignments in it.
We furthermore show that our approach performs better than the standard tomographic method to reconstruct features of input states \textit{using the same hardware} --- which in this case consists of performing shadow tomography on the effective measurement describing the overall apparatus~\cite{innocenti2023shadow,acharya2021informationally,nguyen2022optimising}.
The reason we decide to benchmark against the same hardware is that we wish to demonstrate the viability of our QELM strategy in general experimental measurement scenarios. To this end, one thus needs to directly compare the accuracy obtained using QELMs against alternative methods in each different experimental configuration.

To further highlight the capabilities and advantages of our formal approach, we demonstrate that models trained using only separable states are seamlessly capable of estimating entanglement features of previously unseen entangled states, achieving \textit{out-of-distribution generalization}~\cite{caro2023Outofdistribution,pereira2025Outofdistribution}.
This is in stark contrast to most ML-based approaches, which need a sufficiently representative training dataset to function.
Besides providing, to the best of our knowledge, the first experimental implementation of a single-setting state-estimation architecture that is model-independent and allows to witness nonclassical features of input states, our experiment also marks a significant step forward from previously reported reconstructions of single-qubit properties~\cite{suprano2024experimental}.

\begin{figure*}[ht]
    \centering \includegraphics[width=0.8\linewidth]{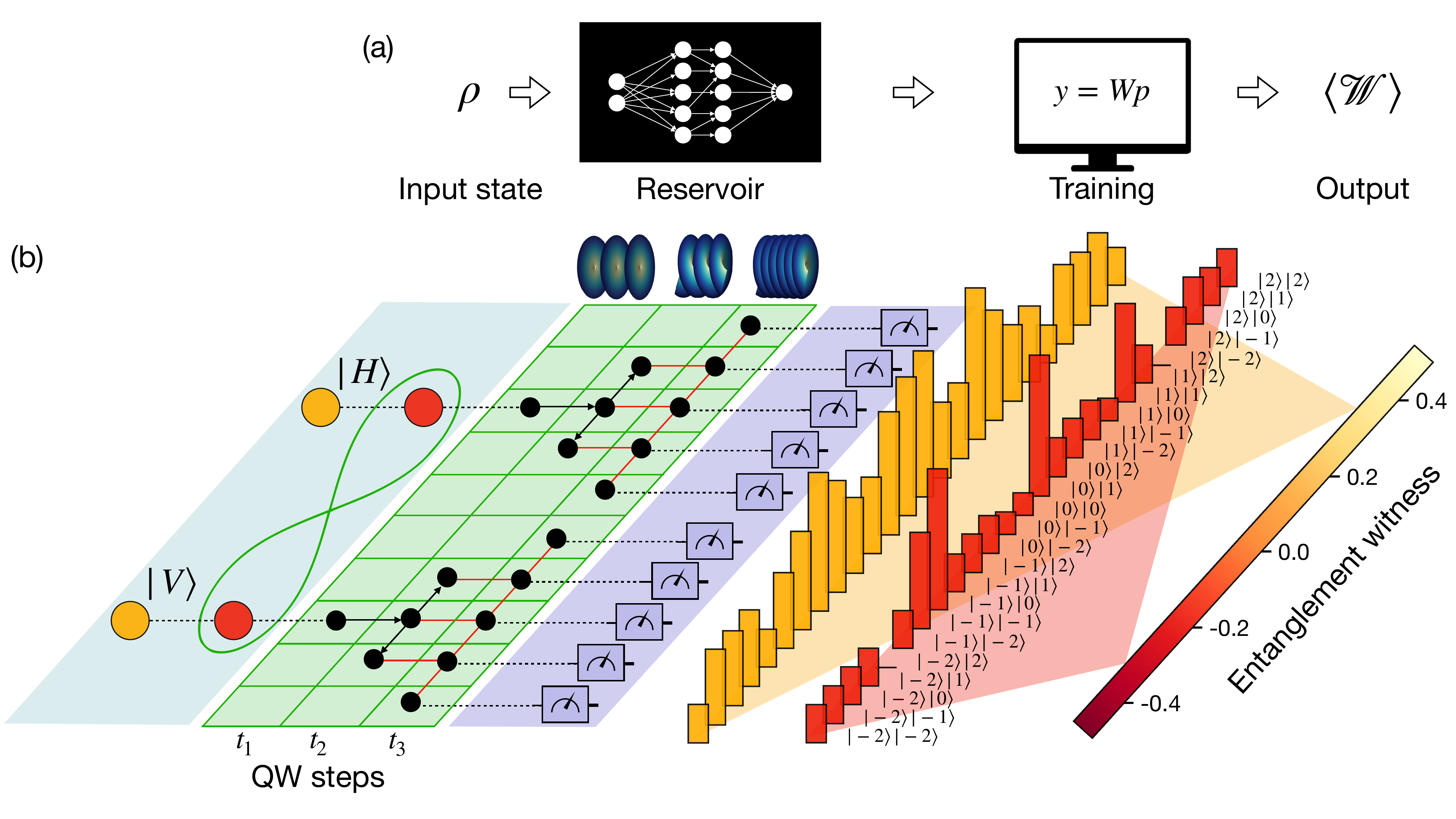}
    \caption{\textbf{Schematic overview of the QELM experiment.} \textbf{(a)} The protocol describes the evolution of an input state through a fixed and unknown quantum channel called reservoir, which maps the state to a higher dimensional Hilbert space. Here, through projective measurements in a single set configuration, the output probability distribution $\{ p \}$ of the state over the basis 
    $\bs\mu = \{\mu_b\}_{b=1}^{N_{\rm out}}$ is retrieved. These probabilities are used in the final training stage to reconstruct the expectation value of an observable on the state. \textbf{(b)} In the experimental realization, entangled and separable quantum states encoded in the polarization degree of freedom of photon pairs evolve through the reservoir dynamics implemented by a double quantum walk configuration. Through the evolution, the 4-dimensional input space is enlarged into the 25-dimensional space of the orbital angular momentum. By performing projective measurements on the computational basis of the latter, we obtain the output outcome probabilities on which the QELM model is trained to reconstruct a target entanglement witness.}
    \label{fig:conceptual_protocol}
\end{figure*}

\section{Results}
\subsection{QELM-based entanglement witnessing}

\parTitle{Overview of QELMs}
QELMs offer a streamlined approach to quantum state estimation by leveraging uncharacterized but fixed quantum dynamics~\cite{innocenti2023potential,suprano2024experimental,vetrano2025state}.
In this framework, input states evolve through a quantum channel that acts as a “reservoir” by dispersing the information into a larger Hilbert space prior to measurement.
A linear readout layer is then trained on a set of pre-characterized states to recover the target features from the measurement data.
In doing so, QELMs automatically learn the optimal mapping from measurement outcomes to desired features, thereby eliminating the need to fully characterize the quantum channel or measurement apparatus.
Moreover, by restricting the post-processing to linear operations, the method avoids overfitting when the target features are linear functions of the input density matrices—as is the case for the expectation value of any observable. This robustness arises from the fact that, quantum mechanically, output probabilities are inherently linear functions of the input states.

\parTitle{Formal description}
Formally, a QELM is implemented via a ``reservoir dynamic'' comprising a quantum channel $\Phi$ and a positive operator-valued measure (POVM) $\bs\mu\equiv\{\mu_b\}_{b=1}^{N_{\rm out}}$, where $N_{\rm out}$ denotes the number of measurement outcomes. 
The channel $\Phi$ describes the evolution of each input state, while the POVM describes how the output states are measured.
The training objective is to determine a linear operator $W$ that, when applied to the measurement statistics of an input state $\rho$, extracts its desired features. In the testing stage, the output probabilities $p_b(\rho)\equiv \Trace(\mu_b\Phi(\rho))$ are estimated from a finite statistical sample, yielding estimates $\hat p_b$, which are then combined linearly as $\sum_b W_b\hat p_b$ to produce the final output.
Importantly, this protocol does not require prior knowledge of $\Phi$ or $\bs{\mu}$; their effective action is automatically learned during training.

\parTitle{Training}
The operator $W$ is obtained via a supervised training procedure using a set of known training states $\{\rho_k^{\rm tr}\}_{k=1}^{N_{\rm tr}}$.
For each training state, $N$ copies are generated, evolved, and measured.
The resulting data is organized into a \textit{probability matrix} $\hat{P}_N$, which converges --- in the $N\to\infty$ limit --- to the exact probability matrix $P_{b,k}\equiv\Trace(\mu_b \Phi(\rho_k^{\rm tr}))$.
If the objective is to estimate the expectation value of an observable $\calO$, one solves for $W$ the linear system
\begin{equation}\label{eq:training_linear_system}
    W \hat P_N = M_{\calO},
\end{equation}
where $M_{\calO}$ is a row vector whose $k$-th element is the expectation value $\Trace(\calO\rho_k^{\rm tr})$.
When training for multiple observables $(\calO_j)_{j=1}^{N_{\rm obs}}$, $M_{\calO}$ becomes an $N_{\rm obs}\times N_{\rm tr}$ matrix with entries
$(M_{\cal O})_{j,k}\equiv \Trace(\calO_j\rho_k^{\rm tr})$,
and $W$ correspondingly a $N_{\rm obs}\times N_{\rm out}$ matrix.
In this context, the training states are assumed to be known. In practice this assumption amounts to assuming a high degree of control of the input state preparation stage of the experiment. Imperfections in state preparation will result in imperfections in the estimation.
A straightforward solution to this linear regression problem is provided by the ordinary least squares estimator computed via the pseudoinverse
\begin{equation}
    W = M_{\cal O} \hat P_N^T (\hat P_N \hat P_N^T)^{-1}.
\end{equation}
Alternative approaches, such as ridge regression or gradient-descent-based methods, can be employed when the system is ill-conditioned --- a scenario that may arise with many outcomes and high training statistics~\cite{innocenti2023potential}.
In our experiments, the pseudoinverse method proved optimal, and we thus assume henceforth that $W$ is computed via the pseudoinverse.
The solvability of Eq. (\ref{eq:training_linear_system}) is guaranteed provided that the \textit{effective POVM} $\tilde\mu_b\equiv \Phi^\dagger(\mu_b)$ is informationally complete, the training statistics are sufficiently large, and the training states span linearly the full state space~\cite{innocenti2023potential}.
Under these conditions, the system can be solved for any target observable $\calO$, enabling the QELM to recover arbitrary information about the measured states. For further details on the training process, we refer to section S1 of the Supplementary Information.

\parTitle{Connection with shadow tomography}
QELMs are closely related to shadow tomography methods, whose core idea is that, given suitable measurement schemes, one can construct \textit{shadow estimators} that accurately estimate target features with relatively few measurements~\cite{huang2020predicting,innocenti2023shadow}.
Effectively, training a QELM produces a shadow estimator tailored to the specific apparatus without requiring explicit knowledge of its internal workings:
in the limit of large training statistics, the linear operator $W$ converges to the estimator employed in shadow tomography~\cite{innocenti2023potential,innocenti2023shadow}.
In other words, QELMs exchange the need for a full characterization of the measurement apparatus --- required in shadow tomography --- for the simpler task of characterizing a restricted set of input states. This shift is particularly advantageous when accurately preparing the input states is easier than accurately modeling the entire experimental apparatus.

\parTitle{Optical implementation}
In our experiment, the quantum channel is realized as an isometric evolution implemented via a two-photon quantum walk (QW) in polarization and OAM.
Each photon evolves independently through a QW apparatus composed of a sequence of polarization waveplates and {q-plates}~\cite{innocenti2017quantum,giordani2019experimental,giordani2020machine,suprano2021dynamical,suprano2021enhanced,giordani2021entanglement,suprano2024experimental}, more details can be found in the Methods section. This design embeds the information encoded in the polarization degree of freedom into the larger OAM space. After projecting the polarization onto a fixed direction, a projective measurement is performed over the OAM computational basis.
In this configuration, the OAM serves as a reservoir --- namely, an ancillary degree of freedom that enables the implementation of a single-setting, informationally complete measurement of the polarization degrees of freedom.
Formally, the {evolution of the initial state $\rho$} is described by the completely positive map
\begin{equation}
        \Phi(\rho)
        =\bra{\eta_p} U(\rho\otimes \ketbra 0) U^\dagger\ket{\eta_p},
\label{eq:QELM_evolution}
\end{equation}
where $U\equiv U_1\otimes U_2$ is the overall unitary evolution implemented by the apparatus, $U_1,U_2$ correspond to the two independent quantum walks, $\ket{\eta_p}$ is the projection onto the photon polarization, and $\ket0$ denotes the initial OAM state of the photons. 
To test the performance of the reservoir in different scenarios, we repeat the experiment with multiple quantum walk configurations, by changing the rotation angles of the waveplates.
Owing to imperfections and thermal fluctuations, the actual apparatus deviates from this idealized model.
However, it is important to stress that this idealized model does not enter the training stage, meaning that the performance of the QELM remains completely agnostic to it.

\parTitle{Estimating entanglement witnesses}
Once the QELM is trained on a complete set of states—that is, a set of density matrices spanning the entire state space—it can be used to estimate arbitrary observables for new, previously unseen quantum states.
Crucially, even if this training set consists only of pure product states, the QELM can still capture entanglement features of unseen entangled states, as first remarked in~\cite{krisnanda2023quantum}.
To demonstrate this capability, we use the QELM to estimate the expectation values of \textit{entanglement witnesses}.
An entanglement witness $\calW$ is an observable whose expectation value is guaranteed to be non-negative for all separable states, but negative for at least one entangled state~\cite{guhne2009entanglement}.
Hence, measuring $\Trace(\calW \rho)<0$ certifies that the state $\rho$ is indeed entangled.
A standard way to construct entanglement witnesses is by defining the observable~\cite{guhne2009entanglement,barbieri2003detection}
\begin{equation}
    \mathcal{W} = \alpha I - \ketbra\psi,
\end{equation}
where $I$ is the identity operator, $\vert \psi \rangle$ is a {\it target} entangled state, and
the parameter $\alpha$ is given as
\begin{equation}
    \alpha = \max_{\rho\,\text{separable}}\text{Tr}(\rho \vert \psi \rangle \langle \psi \vert),
\end{equation}
thus representing the maximum overlap between the target state and separable ones. A depiction of the formal approach entailed by our QELM-based method is given in Fig.~\ref{fig:conceptual_protocol}.

\subsection{Experimental Implementation}

\begin{figure*}[htb!]
    \centering \includegraphics[width=0.99\linewidth]{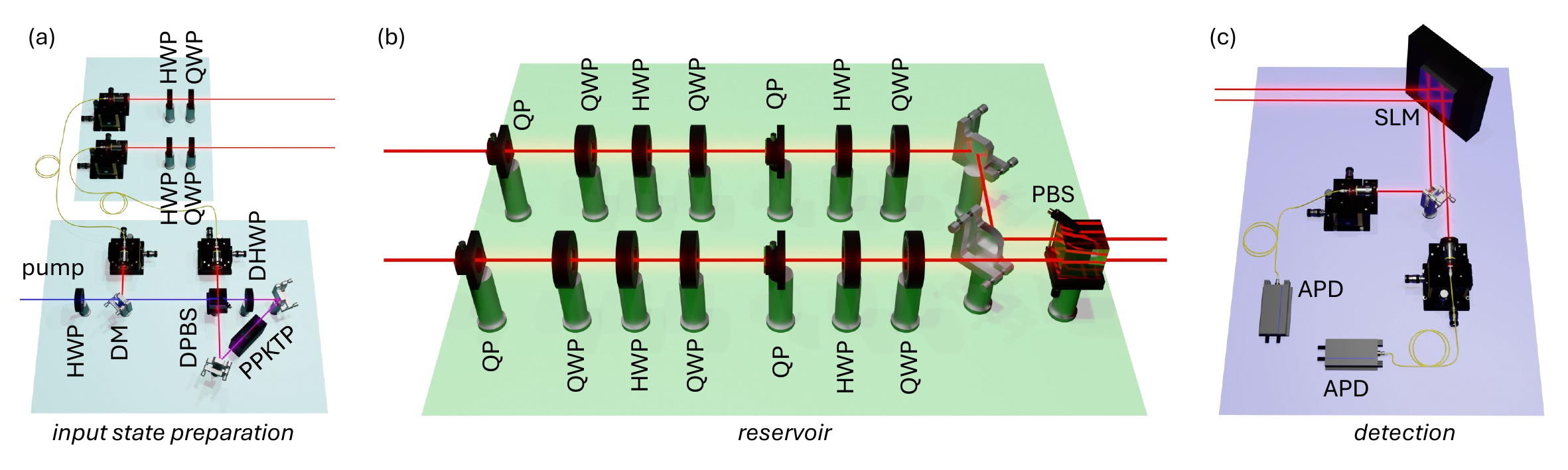}
    \caption{
    \textbf{Experimental Setup.}
    \textbf{(a)}
    \textit{Input state preparation} ---
    Photon pairs are generated by spontaneous parametric down-conversion in a Type-II periodically poled potassium titanyl phosphate (PPKTP) crystal enclosed in a Sagnac interferometer.
    In the Sagnac interferometer, a dual-wavelength half-wave plate (DHWP) allows the nonlinear process to happen in each arm while compensating for the propagation delay acquired inside the crystal by the orthogonally polarized photons.
    The generated photons are separated by a dual-wavelength polarizing beam splitter (DPBS) while the pump laser is separated by a dichroic mirror (DM).
    After generation, each photon enters a layer consisting of a half-wave plate (HWP) and a quarter-wave plate (QWP) which encodes the input polarization state.
    \textbf{(b)}
    \textit{Reservoir evolution} ---
    After the state preparation, each photon enters an independent discrete-time quantum walk (QW) consisting of a series of HWPs, QWPs, and q-plates (QPs), which transfer the polarization information into the OAM degree of freedom.
    This QW implements the reservoir dynamics needed by the QELM.
    The polarization of the photons exiting the QW is then projected with a HWP, a QWP, and a polarizing beam splitter (PBS).
    \textbf{(c)}
    \textit{Detection} ---
    The final detection stage consists of a projective measurement in the OAM space, realized by a spatial light modulator (SLM) followed by coupling into single-mode fibers.
    This implements a projective measurement in the basis $\{\ket n:\, n=-2,\dots,2\}$.
    Finally, avalanche photodiode detectors (APDs) are used to collect the photons and detect the coincidence counts.
    }
    \label{fig:exp_setup}
\end{figure*}

\parTitle{Input states preparation}
A schematic representation of the experimental apparatus is reported in Fig.~\ref{fig:exp_setup}.
The input state preparation stage involves the spontaneous parametric down-conversion (SPDC) photon-pair source, with a periodically poled potassium titanyl phosphate (PPKTP) crystal in a Sagnac configuration. This architecture enables the generation of stable polarization-entangled states. By adjusting the pump polarization, we can switch between generating separable and maximally entangled states.
We will refer to these states as ``reference states'', to distinguish them from the input states that are generated from them.
The reference entangled state is in all cases $\ket*{\Psi^+}=\frac{1}{\sqrt2}(\ket{HV}+\ket{VH})$.
After the generation, the two photons are separated and sent to the input layer of the setup, where each one is manipulated using a half-wave plate (HWP) and a quarter-wave plate (QWP),
with orientations chosen at random and defining the input polarization states entering the reservoir. Overall, by rotating also the waveplate in the pump beam, two distinct sets of random two-qubit polarization states for training and testing the QELM are generated, one consisting of product states and the other of maximally entangled states.

\parTitle{Reservoir dynamic and measurement}
Once generated, the two-photon states are fed into the core computational layer, where the reservoir dynamics are realized by two parallel QW evolutions on both polarization and OAM degrees of freedom of the photons~\cite{innocenti2017quantum,giordani2019experimental} (see~\cref{fig:exp_setup} and the Methods for further details on the experimental implementation).
The resulting double QW evolution generates output states spanning a 25-dimensional walker space, given by the OAM values $\{\ket{n}$, $n \in [-2,2]\}$ for each photon.
After the reservoir evolution, each photon's polarization is projected using a combination of waveplates and a polarizing beam splitter (PBS). Next, a spatial light modulator (SLM) combined with the coupling to single-mode fibers performs projective measurements in the two-photon OAM space. Finally, the photons are detected by avalanche photodiodes (APDs).
Thanks to the dimensionality increase induced by the QWs, this final projective OAM measurement results in a non-projective, single-setting, informationally-complete POVM on the two-photon input.
Repeating this procedure multiple times for each input state, we obtain the dataset used to train and test the QELM.

\begin{figure*}[t!]
    \centering \includegraphics[width=0.95\linewidth]{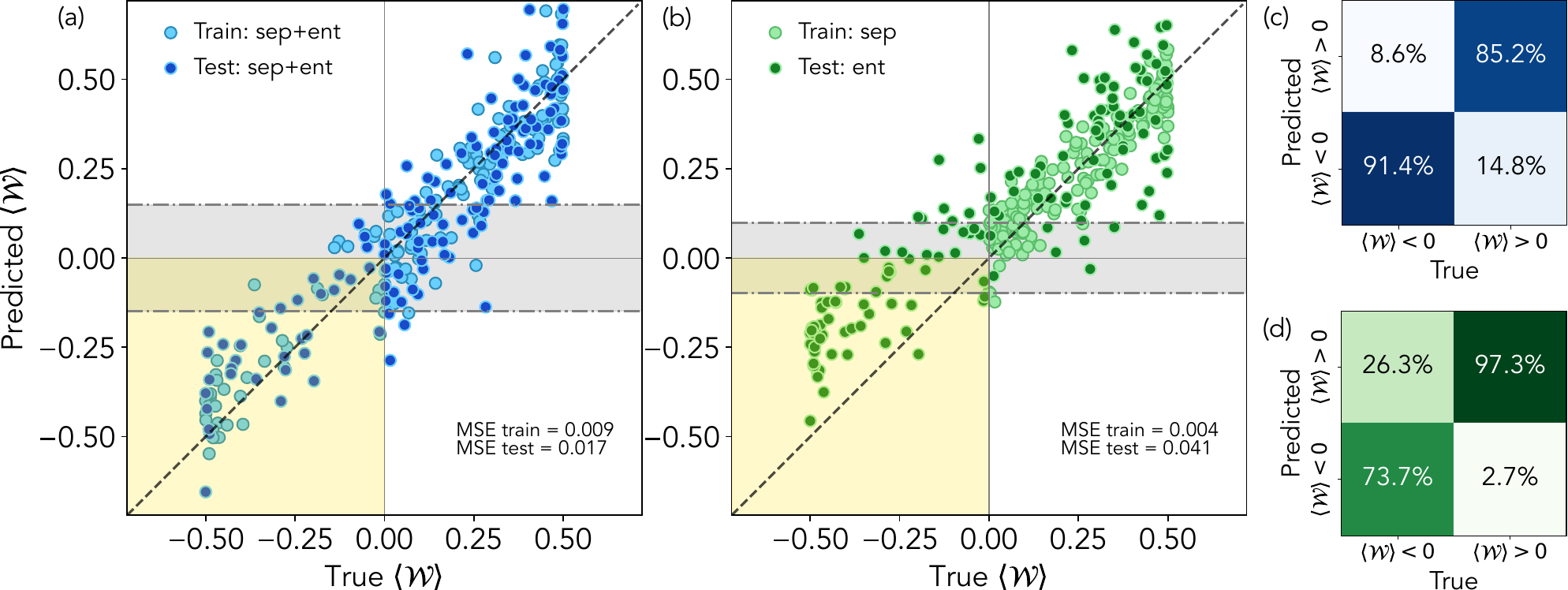}
    \caption{
    \textbf{Performance of entanglement witness estimation}.
    Predicted versus true values of the witness $\mathcal{W}$ for the \E1 scenario under different training configurations.
    In panel \textbf{(a)}, both training and test datasets contain separable \textit{and} entangled states, while in panel \textbf{(b)}, the training set includes only separable states and the test set only entangled ones.
    These results showcase how the model's linearity allows it to train on only states with $\expval{\calW}>0$ and still accurately predict previously unseen states with $\expval{\calW}<0$ (yellow area in the plot).
    The shaded grey region in each plot represents the estimation error, defined as the square root of the training MSE.
    Finally, panels \textbf{(c)} and \textbf{(d)} report the confusion matrices showing the accuracy in correctly identifying positive and negative values of $\expval\calW$, corresponding to the data in \textbf{(a)} and \textbf{(b)}, respectively.
    }
    \label{fig:results}
\end{figure*}

\parTitle{Summary of experimental settings}To benchmark the performance of our platform we performed experiments under three configurations, labeled E1, E2, E3. 
Each configuration is characterized by different QWPs and HWPs angles specifying the reservoir's unitary, and by the method used to prepare the input states. Specifically:
\begin{enumerate}
    \item E1 uses wave-plates angles that were numerically optimized to minimize the average reconstruction mean-squared error (MSE), similarly to how is done in [52].
    \item E2 is derived from E1 by swapping two of the optimized angles, perturbing the reservoir dynamics.
    \item E3 uses a fully random set of waveplate angles, sampled uniformly.
\end{enumerate}
These choices were made to benchmark the hardware performance across ideal (E1), perturbed (E2), and ``random'' (E3) scenarios.
The reference separable state --- that is, the photons state that exits the photons generation and enters the state preparation layer when using separable states --- is $\ket*{\Psi_R}=\ket{VV}$ for \E1 and $\ket*{\Psi_R}=\ket{VH}$ for \E2 and \E3.
These choices are made to test the robustness of the configuration under different conditions, input states, and reservoir dynamics.
In all cases, the same angles are used in both QWs, so that every injected state has the form $(U_{\rm in}\otimes U_{\rm in})\ket*{\Psi_R}$, with $U_{\rm in}$ the single-photon reservoir evolution.
In each configurations we test an equal number of separable and entangled states, namely $150$ for \E1 and \E3, and $83$ for \E2. This difference in the statistics used in \E1 vs \E2 and \E3 is simply due to experimental issues during the data acquisition process; we expect this difference to bear no significant impact on the reported results, and we only report it for completeness.
The preparation angles are sampled uniformly at random in all cases. Additional experimental details can be found in sections S2 and S3 of the Supplementary Information.

\parTitle{Entanglement witness performance with mixed training}
We consider here the task of reconstructing the expectation value of the entanglement witness $\calW\equiv \frac12(I-2\ketbra*{\Phi^+})$, with $\ket*{\Phi^+}\equiv\frac1{\sqrt2}(\ket{HH}+\ket{VV})$, when training the QELM with both separable and entangled states. Notably, our protocol works independently from the choice of the specific witness as long as the corresponding operator is linear with respect to the input state \cite{innocenti2023potential}.
To assess reconstruction performances, we report the mean-squared error (MSE) on both training and test states. The MSE is here defined as the average squared difference between predicted and true $\langle\calW\rangle$, for each state. This directly quantifies the absolute estimation error, which is particularly relevant in our setting where the dominant contribution to the total error arises from stochastic noise that is independent from the magnitude of the observable being estimated
\Cref{fig:results} reports the results for the \E1 configuration, in which training and testing use two sets, each comprising 150 separable and 150 entangled states.
\Cref{fig:results}-(a) demonstrates a strong agreement between reconstructed and true expectation values. The MSEs averaged over the training and test states are $\mathrm{MSE}_{\rm train}\simeq 0.009$ for training, $\mathrm{MSE}_{\rm test}\simeq 0.017$ for testing.
From the perspective of certifying entanglement, we quantify the apparatus accuracy as the fraction of entangled states with $\langle\calW\rangle<0$, for which the model also correctly predicts $\langle\calW\rangle<0$.
These correspond to the points in the shaded lower-left region in~\cref{fig:results}-(a).
As shown in the confusion matrix in~\cref{fig:results}-(c), entanglement is correctly certified for $91.4\%$ of these states.
These estimates are subject to stochastic errors arising, among other factors, from finite sampling statistics.
It is therefore sensible to consider a state as reliably certified as entangled only if the estimated witness value is more negative than a threshold determined by the training MSE.
Taking this into account --- and using a threshold of three standard deviations, where the standard deviation is computed as the square root of the training MSE --- we find that $37.1\%$ of entangled states remain certifiably entangled even within this elevated error tolerance.
These numbers depend weakly on the specific subset of random training states chosen.
Averaging these errors over random training instances we get $\overline{\mathrm{MSE}}_{\rm train}= 0.010 \pm 0.001$ and $\overline{\mathrm{MSE}}_{\rm test}= 0.015 \pm 0.001$, with a number of certified entangled states equal to $(89 \pm 4)\%$. The reported errors represent one standard deviation.
These results showcase the good estimation performances of the QELM strategy even in relatively noisy conditions, and without the protocol requiring knowledge of the underlying dynamic.

\parTitle{Generalization performances}
Another notable consequence of QELMs restricting to linear post-processing is the model's amenability to domain generalization, particularly in the context of entanglement witnessing.
Because training a QELM enables it to learn how to estimate target observables, and because separable states alone span the space of all states, it follows that a QELM can be trained exclusively on separable states yet still perform well on previously unseen entangled states.
To demonstrate this, we present in~\cref{fig:results}-(b) the results obtained when the training dataset contains all 150 separable states, while the test dataset contains all 150 entangled ones.
We again observe strong agreement between the true and predicted values of $\langle\calW\rangle$, and that many states are reliably identified as entangled --- even though the model was never trained on any state corresponding to negative $\langle \calW\rangle$.
The training and test MSEs are $\mathrm{MSE}_{\rm train}\simeq 0.004$ and $\mathrm{MSE}_{\rm test}\simeq 0.04$.
The fraction of correctly certified entangled states, as reported in~\cref{fig:results}-(d), is $73.7\%$.
Accounting for stochastic errors, $32.9\%$ of entangled states remain certified within three standard deviations.

\begin{figure*}
    \centering
    \includegraphics[width=\linewidth]{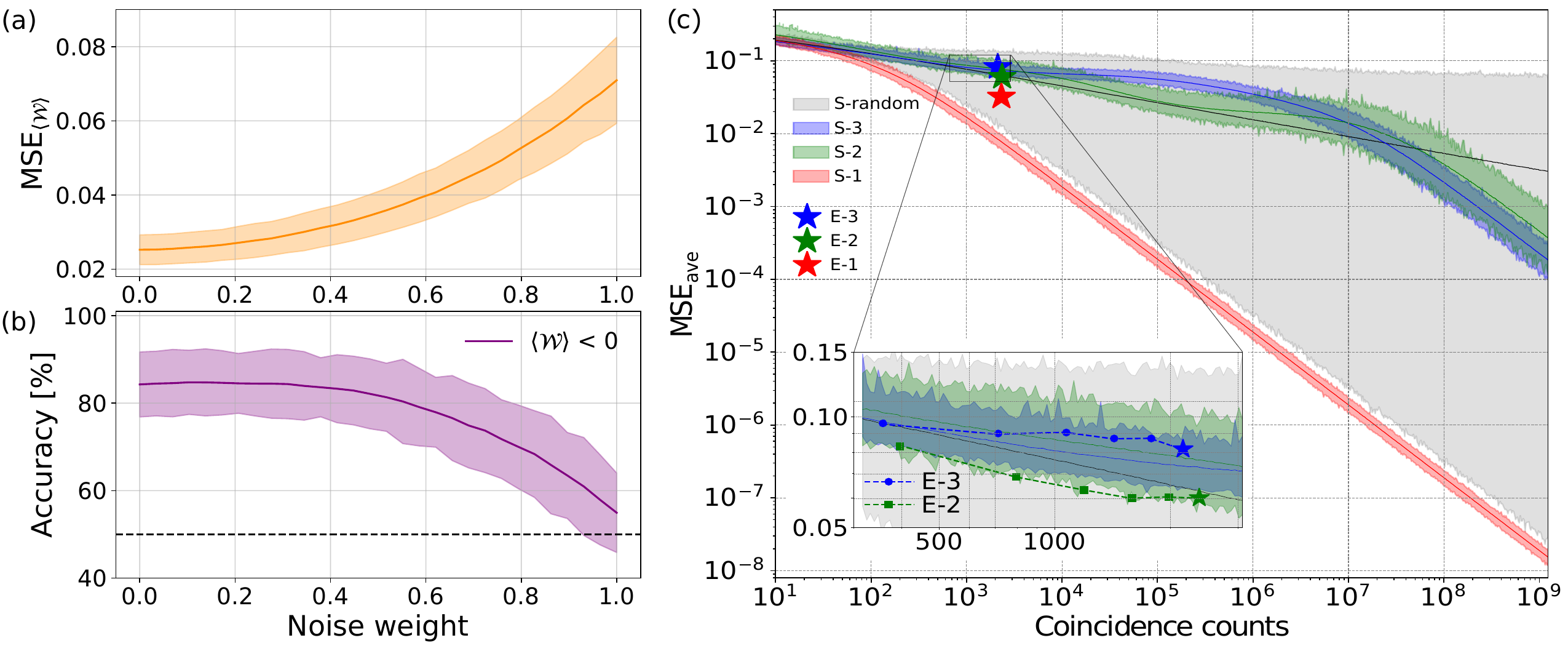}
    \caption{
    \textbf{Dependence of estimation performances on noise and choice of reservoir.}
    Panels \textbf{(a)} and \textbf{(b)} report the robustness of the witness estimate $\langle\calW\rangle$ under noise, where the entangled reference state $\rho_{\rm ent}$ is replaced by statistical mixtures of the form $\rho_{noise} = (1 - p) \rho_{\mathrm{ent}}  + p \rho_{\mathrm{sep}}$, where $p\in[0,1]$ is the noise parameter. Here, the studies are performed in the experimental scenario \E2. This scenario is not the optimal one but is obtained for a perturbed reservoir dynamic, swapping the QWP angles.
    In panel \textbf{(a)}, the solid curve shows the average MSE on the test set as a function of $p$, and the shaded regions represent the standard deviations obtained by repeating the procedure with different random training instances. Each instance contains 150 training states and 150 test states.
    Panel \textbf{(b)} reports the corresponding accuracy in identifying negative values of the $\langle\calW\rangle$, the dotted line indicates the $50\%$ accuracy value.
    Panel \textbf{(c)} compares experimental performances (\E1, \E2, \E3) with numerical simulations ($\mathsf S1$, $\mathsf S2$, $\mathsf S3$).
    Stars indicate the experimental MSEs, while solid lines represent the corresponding numerical results under the same conditions and color coding.
    In all cases, training is performed on separable states and testing on entangled states; the MSE is averaged over the possible target observables and plotted against the collected statistics (fixed as the same in both training and test).
    The shaded regions around each simulation curve indicate one standard deviation across sampling instances, while the boundaries of the gray shaded region mark the minimum and maximum averaged MSE obtained when training is performed with random reservoir configurations ($\mathsf S$-random).
    The inset details the behavior of the experimental results for \E2 and \E3, showing the consistency between numerical simulations and experimental data across varying statistics.
    }
    \label{fig:noise}
\end{figure*}

\parTitle{Effect of input states mislabeling}
To evaluate the noise resilience of the model, we constructed a new dataset of non-maximally-entangled states by taking a weighted combination of the experimentally measured OAM distributions for entangled and separable states.
This amounts to using as reference states for the entangled systems statistical mixtures of the form $\mathbf{\rho}_{\mathrm{noise}} = (1 - p) \mathbf{\rho}_{\mathrm{ent}} + p \mathbf{\rho}_{\mathrm{sep}}$, with $p$ a noise parameter, $\rho_{\rm ent}$ the reference entangled state, and $\rho_{\rm sep}$ the reference separable state.
In any case, each of these mixed states was labeled with the expected value computed for the maximally entangled state. The study was performed in the experimental scenario \E2 and 
Fig.~\ref{fig:noise}-(a,b) illustrate the impact of noise on the model's performance.
In particular, the average MSE and the accuracy in identifying negative entanglement witness values as functions of $p$ are reported, respectively in the two panels.
These results indicate that the model is highly resilient to noise, with performance remaining robust for noise levels up to $p = 0.5$.

\parTitle{Dependence of performances on reservoir configuration}
In Fig.~\ref{fig:noise}-(c), we compare the performance of the three configurations in the numerical simulations and the experimental implementations.
The data illustrates that different reservoir configurations can yield different estimation accuracies under the same training statistics.
This is due to different configurations corresponding to different effective POVMs in the Heisenberg picture, and thus naturally different estimation performances~\cite{innocenti2023potential}.
Although our double-QW apparatus employs a relatively simple evolution --- far from the ideal random unitary scenario associated with optimal performance --- we still observe, at least in numerical simulations, the expected inverse proportionality between the average MSE and the number of collected samples $N$.
This emerges after an initial settling regime in which the test statistics cannot be fully exploited due to insufficient training statistics: until the estimator $W$ is sufficiently unbiased, the MSE decreases at a slower rate than $1/N$.
For sufficiently large training statistics, we recover the expected behavior where the test MSE decreases as $\beta/N$, with $\beta$ a proportionality constant that depends on the specifics of the measurement setting.
Different reservoir configurations spread the encoded information among outcomes to varying degrees, which in turn influences how readily that information can be retrieved. As a result, certain choices of angles --- in particular, those used in the \E1 configuration --- are sufficient to reach the $1/N$ regime with the available experimental statistics.
In contrast, the configurations \E2 and \E3 require additional statistics before reaching the same performance level. In principle, this regime could be achieved experimentally by increasing the data acquisition
time and, therefore, the number of coincidence counts for each state. For pratical reasons we have set the acquistion time to complete a full data acquisition within 10 hours having an average overall signal of $\sim$ 3000 coincidence counts per state. Note that the apparently larger discrepancy between \E1 experimental and simulated data is an artifact of the log-log scale: the discrepancy is in all cases found to have the same order of magnitude of $\sim 10^{-2}$.

\parTitle{Benchmark with shadow tomography}
To benchmark the performance of our QELM-based approach, we compare it against shadow tomography~\cite{huang2020predicting,elben2023randomized,gebhart2023learning,innocenti2023shadow}, a general methodology for computing the optimal unbiased estimator of target observables --- that is, the unbiased estimator with the smallest average variance ---  in fixed-measurement scenarios. Because both approaches aim to efficiently extract properties of input states without requiring a full tomographic reconstruction, shadow tomography is the natural benchmark for our apparatus.
For reservoir configuration \E1, and the same task as~\cref{fig:results}, shadow tomography yields an MSE of approximately \(0.015\) for separable states and \(0.072\) for entangled states. In contrast our QELM, trained solely on separable states, reduces these MSEs to \(0.002\) and \(0.041\), respectively. Doing the training on both separable and entangled states gives the even smaller MSEs \(0.009\) and \(0.017\). Moreover, the reported shadow-tomography MSEs represent lower bounds, since shadow tomography requires knowledge of the input statistics \(N\), which is unavailable here. These findings clearly highlight the advantages of using a QELM-based approach for state-property estimation.
In principle, one \textit{could} build a much more detailed physical model of the apparatus and, with it, push the shadow-tomography performance beyond that of the QELM. In practice, however, constructing and validating such a model is very time-consuming and resource-intensive, and very susceptible to the specifics of the hardware. By contrast, our model-agnostic QELM pipeline avoids this overhead: it learns directly from experimental data and therefore adapts naturally to modifications in the setup while still delivering strong performance.
For further details on this analysis we refer to section S4 of the Supplementary Information.

\section{Discussion}

We have experimentally demonstrated the capability of quantum reservoir computing, specifically in the QELM form, to reconstruct an entanglement witness for bipartite states of a two-photon system, achieving performance on par with methods that rely on careful calibration and prior knowledge of the apparatus. Crucially, our approach is platform-independent: the QELM paradigm can be straightforwardly adapted to different experimental settings, thanks to its ability to adapt to measurement outcomes on-the-fly without explicit modeling of the underlying dynamics.

In particular, our photonic implementation leverages the orbital angular momentum degree of freedom to realize a compact, single-setting measurement scheme that bypasses many practical issues of conventional multi-setting protocols, such as the need for frequent device-reconfiguration or stringent calibration requirements. As the QELM training procedure reduces to linear regression, we also circumvent interpretability challenges and overfitting issues commonly encountered in more complex machine learning architectures.

A key advantage of our method is its robustness: by demonstrating that training solely on separable states is sufficient to accurately witness the entanglement of new, previously unseen input states, we highlight how the QELM effectively \emph{self-calibrates} to the experiment. Moreover, our findings reveal that drifts in the experimental setup can leave the final witness reconstruction largely unaffected, thus highlighting the resilience of such a technique.

Looking ahead, the QELM principles deployed here can be applied to more general quantum state estimation scenarios, paving the way for improved property-estimation tasks in complex photonic platforms and beyond. Equally remarkable is the prospect of \emph{transfer learning} demonstrations, where QELMs trained on fully classical data (e.g. coherent-states) can later be applied to certify nonclassical properties of unknown quantum states, with no additional overhead. In virtue of its simplicity, flexibility and generality, this method has thus the potential to become a valuable tool for the characterization of quantum technologies.

\section{Materials and Methods}

\parTitle{Quantum walks in the angular momentum} The quantum walk evolution in the angular momentum of single photons is realized by encoding the \textit{coin} and \textit{walker} states, respectively, in their polarization and OAM degrees of freedom~\cite{giordani2019experimental}. 
Each QW step consists of a rotation of the internal coin state, implemented by three waveplates, followed by a controlled shift on the walker position, realized by the QP . The latter is a device made of a birefringent material designed to have a non-uniform pattern for its optical axis, which is engineered to enable the conditional change of the light OAM in function of its polarization ~\cite{marrucci2006optical}.
Formally, each coin operation is given by $C(\zeta,\theta,\phi) = \on{QWP}(\phi)\on{HWP}(\theta)\on{QWP}(\zeta)$, where $\{\zeta, \theta, \phi\}$ are the waveplates rotation angles; the controlled-shift operator $S(\alpha,\delta)$ instead depends on the tuning parameter $\delta$, which governs the coupling efficiency of the QP, and $\alpha$, the initial angle of the device optical axis respect to the horizontal one.
Explicitly, using the circular polarization basis for the polarization, we have that
\begin{equation}
\begin{aligned}
    C(\zeta,\theta,\phi) = \begin{pmatrix}
         e^{-i(\zeta{-}\phi)}\cos\eta & e^{i(\zeta{+}\phi)}\sin\eta\\
        -e^{-i(\zeta{+}\phi)}\sin\eta & 
        e^{i(\zeta{-}\phi)}\cos\eta
    \end{pmatrix},  
\end{aligned}
\end{equation}
\begin{equation}
\begin{aligned}
        S(\alpha,\delta) =
        \sum_{n=-N+1}^{N-1}\cos\frac{\delta}{2}\left(\ket{L,n}\bra{L,n}+\ket{R,n}\bra{R,n}\right)\\
        +i \sin\frac{\delta}{2}\left(e^{2i\alpha}\ket{L,n}\bra{R,n{+}1}
        +
        \mathrm{h.c.}
        \right),
\end{aligned} 
\end{equation}
where $\eta = \zeta - 2 \theta + \phi$, and $\ket{R}$($\ket{L}$) and $\ket{n}$ denote the right (left) circular polarization state and the OAM eigenstates, respectively.
Adjusting $\delta$ allows partial controlled-shift operations, which serves to further enlarge the accessible OAM space by employing the unchanged component of the OAM~\cite{suprano2024experimental}.
In particular, we implement a two-step evolution in each of the two QWs
, so that the overall unitary is $U=U_1\otimes U_2$, with
\begin{equation}
    U_k = S(\alpha^k_2, \pi) C(\zeta^k, \theta^k, \phi^k) S(\alpha^k_1, \pi/2),
    \, k\in\{1,2\}.
\label{eq:total_U}
\end{equation}

The parameters $\alpha_1^1=19^\circ$, $\alpha_2^1=77^\circ$, $\alpha_1^2=336^\circ$, $\alpha_2^2=163^\circ$ are fixed by the QP fabrication process.
After the evolution through this apparatus, each photon is in a superposition of the five OAM states $\{\ket{-2},\ket{-1},\ket{0},\ket{1},\ket{2}\}$, leading to a total 25-dimensional output walker space.

\section{Acknowledgments}
\subsection{Funding}
This work was supported from ICSC– Centro Nazionale di Ricerca in High Performance Computing, Big Data and Quantum Computing, funded by European Union– NextGenerationEU. F. S. and V. C. acknowledge support from the project QU-DICE, Fare Ricerca in Italia, Ministero dell'istruzione e del merito, code: R20TRHTSPA.  M. P. acknowledges support from the European Union’s Horizon Europe EIC-Pathfinder project QuCoM (101046973), the Royal Society Wolfson Fellowship (RSWF/R3/183013), the UK EPSRC (EP/T028424/1), the Department for the Economy Northern Ireland under the US-Ireland R\&D Partnership Programme. 
G.M.P., S. L., L. I. And M. P. are supported by the ICSC– Centro Nazionale di Ricerca in High Performance Computing, Big Data and Quantum Computing, funded by European Union– NextGenerationEU, The “National Centre for HPC, Big Data and Quantum Computing (HPC)” (CN00000013) – SPOKE 10 through project HyQELM; L. I. and G. M. P. acknowledge MUR and AWS under project PON Ricerca e Innovazione 2014-2020, ``Calcolo quantistico in dispositivi quantistici rumorosi nel regime di scala intermedia" (NISQ - Noisy, Intermediate-Scale Quantum). A.F., G.M.P. and S.L. are supported by MUR under PRIN Project No. 2022FEXLYB ``Quantum Reservoir Computing" (QuReCo).

\subsection{Author contributions}
D. Z., L. I., S. L., A. S., N. S., T. G., V. C., G. M. P., A. F., F. S., and M. P. conceived the experiment. D.Z., G. M., A. S., R. Di B., N. S., T. G., V. C., and F.S. developed the experimental platform and performed the experiment. D.Z., L. I., G. M., S. L., T. G., V. C., F. S., and M. P.   carried out the data analysis. All authors contributed to discussing the results and to writing the manuscript.
\subsection{Competing interests}
The authors declare that they have no competing interests. 
\subsection{Data and materials availability}
All data needed to evaluate the conclusions in the paper will be available in Zenodo.

\bibliography{arxiv_v2_main}

\clearpage
\onecolumngrid
\appendix
\renewcommand{\thefigure}{S\arabic{figure}}
\renewcommand{\thetable}{S\arabic{table}}
\renewcommand{\theequation}{S\arabic{equation}}
\setcounter{figure}{0}
\setcounter{table}{0}
\setcounter{equation}{0}

\section*{Supplemental Material}

\section{Theoretical background}
\label{sec:theoretical_shit}

\parTitle{General recipe}
Building on a quantum reservoir-based architecture, a QELM uses a training dataset of \emph{a priori} known states, each evolving under an unknown ``reservoir dynamic'', to learn how best to extract information from measurement data on new, previously unseen states. 
We can always model the reservoir dynamic as a quantum channel $\Phi$ followed by a POVM $\bs\mu \equiv \{\mu_1,\ldots,\mu_{N_{\rm out}}\}$, with $N_{\rm out}$ the number of measurement outcomes.
The training dataset comprises pairs $\{(\rho_i^{\rm tr},o_i)\}_{i=1}^{N_{\rm tr}}$, where $\rho_i^{\rm tr}$ is the $i$-th training state and $o_i$ is the target value the QELM should output for $\rho_i^{\rm tr}$.
Alternatively, one can consider as training dataset the pairs $\{(f^{\rm tr}_i,o_i)\}_{i=1}^{N_{\rm tr}}$, with $f^{\rm tr}_i$ the set of frequencies obtained measuring $\rho_i^{\rm tr}$ a set number of times.
Unless otherwise specified, $o_i$ is taken to be an expectation value of one or more observables $\calO^{(j)}$. 
Concretely, either $o_i = \Trace(\rho_i^{\rm tr}\,\calO)$ for some single observable $\calO$, or $\bs o_i = \bigl(\Trace(\rho_i^{\rm tr}\,\calO^{(j)})\bigr)_{j=1}^{N_{\rm obs}}$ for $N_{\rm obs}>1$ observables.
Ultimately, the QELM is trained to produce a linear function $W$, which is then applied to measurement data from any --- possibly previously unseen --- input state to recover the desired expectation values.
In this work, the target observable is taken to be either a tensor product of some pair of Pauli matrices, or an entanglement witness of the form $\calW_{\Psi^{\rm Bell}_i}=\frac I2-\ketbra*{\Psi^{\rm Bell}_i}$, where $\ket*{\Psi^{\rm Bell}_i}\in\{\ket*{\Phi^\pm},\ket*{\Psi^\pm} \}$ are the four canonical Bell states.
For brevity we will refer to these four witnesses as $\calW_i$, $i=1,2,3,4$.

\parTitle{Advantages over alternative methods}
Restricting a QELM to linear post-processing provides several advantages compared to more general machine-learning approaches to quantum state estimation~\cite{ma2018transforming,lu2018separability,palmieri2020experimental,harney2020entanglement,lohani2020machine,gebhart2020neural,giordani2020machine,suprano2021dynamical,suprano2021enhanced,quek2021adaptive,fiderer2021neural,ahmed2021quantum,Scala_2022,hsieh2022direct,schmale2022efficient,Asif2023,greenwood2023machine,cimini2023deep,zia2023regression,melnikov2023quantum,innan2024quantum}. 
In particular, training is simpler, the model is more interpretable, and the risk of overfitting is significantly reduced. 
The key reason is the linearity between input density matrices and outcome probabilities, valid in any quantum system regardless of $\Phi$ and $\bs\mu$. 
This linearity implies that, if the overall measurement is informationally complete, one can recover the input state from outcome probabilities via a linear function --- which is precisely how QELMs operate~\cite{innocenti2023potential}.
If $\Phi$ and $\bs\mu$ were fully characterized, one could use standard shadow tomography~\cite{innocenti2023shadow} 
or full state tomography to estimate observables or reconstruct input states. 
However, these techniques rely on a precise model of the experimental apparatus. 
Any inaccuracy in this modeling degrades performance. 
In contrast, QELMs do not require explicit knowledge of $\Phi$ or $\bs\mu$, instead relying on a pre-characterized (training) set of input states.
This can be advantageous because accurately preparing and modeling a subset of states is often much simpler than modeling the entire apparatus~\cite{suprano2024experimental}.

\parTitle{Effective POVM description}
A convenient way to analyze a QELM is to treat the channel $\Phi$ and measurement $\bs\mu$ as a single \emph{effective POVM} $\bs{\tilde{\mu}} \equiv \{\tilde\mu_1,\ldots,\tilde\mu_{N_{\rm out}}\}$ acting directly on the input states in the Heisenberg picture, where $\tilde\mu_b = \Phi^\dagger(\mu_b)$, and $\Phi^\dagger$ is the adjoint of $\Phi$. 
The outcome probabilities then simply read $p_b(\rho) = \Trace(\tilde\mu_b\,\rho)$. 
From the standpoint of extracting features of the input state, this $\bstildemu$ is the object that more directly affects performances, rather than $\Phi$ and $\bs\mu$ individually.

\parTitle{Training method}
A defining feature of QELMs is their restriction to a \emph{linear} model, meaning measurement data is only post-processed through linear (or affine) operations. 
Once trained, applying a QELM to a new input state $\rho$ involves
(1) measuring a number of copies of $\rho$ to obtain a frequency vector $\bs f(\rho)\in\mathbb{R}^{N_{\rm out}}$ that approximates the true probabilities $\bs p(\rho)=\langle\bstildemu,\rho\rangle$, and
(2) computing $\hat o = W\,\bs p(\rho)$, where $W$ is the matrix obtained from the training.
Finding $W$ involves solving the linear system
    $W\langle\bstildemu,\bs\rho^{\rm tr}\rangle = \langle\bs\calO,\bs\rho^{\rm tr}\rangle,$
where the shorthand $\langle\bstildemu,\bs\rho^{\rm tr}\rangle$ denotes the $N_{\rm out}\times N_{\rm tr}$ matrix whose $j$-th column is the outcome probability vector for the $j$-th training state, and $\langle\bs\calO,\bs\rho^{\rm tr}\rangle$ is the $N_{\rm obs}\times N_{\rm tr}$ matrix containing the corresponding true expectation values of the target observables.
A canonical solution is
    $W = \langle\bs\calO,\bs\rho^{\rm tr}\rangle
    \langle\bstildemu,\bs\rho^{\rm tr}\rangle^+,$
where $(\bullet )^+$ denotes the Moore--Penrose pseudoinverse. 
This solution is guaranteed when $N_{\rm out} \le N_{\rm tr}$ and $\langle\bstildemu,\bs\rho^{\rm tr}\rangle$ is surjective, which corresponds to the effective measurement being informationally complete and there being sufficiently many training states. 
When the number of training states is \textit{strictly} larger than the number of outcomes, $N_{\rm out} < N_{\rm tr}$,
there are multiple possible solutions for $W$. In such cases, the pseudoinverse corresponds to the solution minimizing the Euclidean norm.

\parTitle{Interpretability of the training process}
As shown in [47], the canonical solution for $W$ can be written as
    $W = \langle\bs\calO,\bstildemu^\star\rangle,$
that is, 
$W_{ib} = \Trace(\calO_i\,\tilde\mu_b^\star),$
with $\tilde\mu_b^\star = S^{-1}(\tilde\mu_b)$ the so-called \textit{canonical dual POVM}, and $S$ the quantum map with action $S(X)=\sum_b \tilde\mu_b\Trace(\tilde\mu_b\,X)$ [9, 47].
Thus, training a QELM effectively recovers the dual of the underlying (effective) measurement.

\parTitle{Effects of finite statistics on the training}
The relation between $W$ and $\bstildemu^\star$ holds exactly only when using the exact probabilities in $\langle\bstildemu,\bs\rho^{\rm tr}\rangle$.
In practice, however, one has only access to a finite-sample estimate $\hat{P}_N$, where $N$ is the number of measurement shots per training state. 
Deviations of $\hat{P}_N$ from the true probability matrix lead to a \emph{biased} estimator $W$ for the input state, and thus to biased estimates for any target observable. 
This bias is the counterpart for QELMs of the systematic errors introduced by inaccuracies in the apparatus modelization in other estimation approaches.

\parTitle{Effects of finite statistics on the test}
Once the training has been completed, we have a fixed estimator to be used for appliations.
Explicitly, this estimator is the function $b\mapsto (W_{i,b})_{i=1}^{N_{\rm obs}}$, which assigns the $b$-th column of $W$ to each observation $b$.
Or equivalently stated, if $N_{\rm test}$ measurement shots are taken in testing, with a fixed input state $\rho$, and the observed outcomes are $(b_j)_{j=1}^{N_{\rm test}}$, $b_j\in\{1,...,N_{\rm out}\}$,
then the resulting estimate for the $i$-th target observable is the empirical mean
$\frac{1}{N_{\rm test}}\sum_{j=1}^{ N_{\rm test}} W_{i,b_j}$.
Denoting with $\hat o_{N_{\rm test}}$ this estimator, then by definition $\mathbb{E}[\hat o_{N_{\rm test}}]=\Trace(\rho \calO_i)$, while its variance $\on{Var}[\hat o_{N_{\rm test}}]$ is inversely proportional to $N_{\rm test}$.
Standard statistical bounds like Chebyshev's or Hoeffding's can then be used to derive the statistics required to have a target estimation accuracy with high probability [5]. 

\parTitle{Overfitting}
Overfitting is mitigated because QELMs rely on a linear model, which reflects the physical relationship between input density matrices and measurement probabilities. 
As long as one restricts the target functionals to linear ones (e.g., expectation values of observables) and uses linear regression for training, there is effectively no room for overfitting. 
If the training statistics are insufficient, however, the resulting estimator may carry a bias, leading to systematic estimation errors during testing --- in just the same way as model inaccuracies lead to biases for traditional methods.

\parTitle{Structure of training states}
To reconstruct \emph{any} observable at the testing stage, the training states must span the space of all density matrices. 
Formally, the real linear span of \(\{\rho_i^{\rm tr}\}_{i=1}^{N_{\rm tr}}\) must have dimension \(d^2\), where \(d\) is the dimension of the relevant Hilbert space.
Consequently, \(d^2\) random pure states are in principle sufficient to train a QELM. 
However, as discussed in [47], using more than \(d^2\) states can improve numerical stability in the corresponding linear regression.
Nonetheless, one may successfully train a QELM with less than $d^2$ states if the same restricted subspace of states is also used for testing.
The only requirement is that each testing density matrix can be in the linear span of the training ones.
This enables more efficient training when only a limited set of states is of interest.
Additional details on the training states employed in our experiments are provided in~\cref{sec:reservoir_configurations}.

\section{Experimental implementation details}
\label{sec:implementation_details}

\subsection{Departures from the basic model}
\label{sec:thedeparted}

In this section, we discuss some details of our experimental implementation that deviate from the standard QELM formalization outlined earlier. We examine these details for completeness, although they ultimately have no practical impact on the overall state estimation protocol.

\parTitle{Polarization projection}
Our apparatus requires a polarization projection before OAM measurement because spatial light modulators only operate on photons with a fixed polarization, namely the one oriented as the tilting direction of the liquid crystal molecules in the active area of the device. As a result, the overall mapping from input density matrices to output probabilities is not strictly a quantum channel followed by a POVM. 
Instead, it can be modeled as a non-trace-increasing completely positive map, followed by a projective measurement. 
Equivalently, we can describe the ``effective POVM'' describing the apparatus as a strict subset of a full POVM.
This does not affect our formal framework or training procedure but rather just increases the amount of data (statistics) required for both training and testing [52].

\parTitle{Post-selection measurements}
Because we operate in a post-selection configuration, we do not have access to the total number of input states generated by the source. Instead, we record only the raw photon counts at the output, meaning that the number of copies of each input state is not directly measured, and rely instead on detected events --- specifically, coincidence counts defined as the simultaneous triggering of two APDs within a given time window. Standard estimation methods, including state tomography, depend instead on accurately estimating output probabilities.
Consequently, fluctuations or inaccuracies in the number of input states introduce significant estimation errors and make it impossible to distinguish between apparatus losses and variations in the source generation frequency.
By contrast, QELMs do not require explicit knowledge of the total input count; as long as the overall count rate remains sufficiently stable, the QELM automatically learns the appropriate scaling factor to accurately extract the desired information from the measured counts. We further explore these aspects in~\cref{sec:benchmark_alternative_approaches}.

\parTitle{QELM training with counts vs frequencies}
A potential issue arises if only experimental detection counts, rather than normalized frequencies, are available during training. 
In our photonic implementation, the total number of photon pairs injected into the system is not directly accessible but can only be inferred from a precise calibration of the overall system efficiency. Although we perform coincidence detection to identify successful photon-pair events, in the absence of heralding and due to non-negligible losses and detectors inefficiencies, many generated pairs are lost before detection. Therefore, a more general description relies only on the number of measured events.
The ideal estimator \( W \), derived from normalized frequencies in the limit of infinite training statistics, is given by  
$W =
\langle\bs{\mathcal{O}},\bs{\rho}^{\rm tr}\rangle \langle\bs{\tilde{\mu}},\bs{\rho}^{\rm tr}\rangle^{+}$.
This operator implicitly depends on \( N \) through the term \( \langle\boldsymbol{\tilde{\mu}},\boldsymbol{\rho}^{\rm tr}\rangle \), which contains the output probabilities.
However, if only the detection counts \( N_b \) corresponding to each outcome $b$ are available, we must instead construct a ``counts matrix'', which for large \( N \) approximates
$N\langle\boldsymbol{\tilde{\mu}},\boldsymbol{\rho}^{\rm tr}\rangle.$
Since the computation of \( W \) involves the pseudoinverse of this matrix, the learned estimator obtained from counts ends up being approximately scaled by a factor \( 1/N \).
Consequently, if this estimator is applied to test states corresponding to a different total number of input states \( N' \), the resulting predictions are systematically scaled by a factor \( N'/N \), introducing an unwanted bias.
A practical way to mitigate this issue is to normalize the counts, replacing \( N_b \) with \( N_b / \sum_{b'} N_{b'} \) during both training and testing.  
However, normalization can inject unwanted nonlinearities in the training when the experiment suffers from significant photon losses, which is common in realistic scenarios where \( \sum_{b'} N_{b'} \ll N \).  
Still, under the reasonable assumption that photon losses occur independently of the input state, one can approximate  
$\sum_{b'} N_{b'} \approx \eta N$,
where \( \eta\ll1 \) represents the overall transmission efficiency of the apparatus. Since \( \eta \) is a fixed property of the setup and does not depend on measurement statistics, normalizing the counts does not introduce substantial additional error. Thus, this approach provides a reliable way to ensure consistency between training and testing, even when measurements are performed under different statistical conditions.
In our experiments, training and test statistics are comparable, $N'\simeq N$, and we thus do not observe a substantial difference in the results obtained with and without normalizing experimental counts.
However, this observation tells us that QELMs can be used even in cases where training and test datasets are measured with different statistics.

\parTitle{Implementation of projective OAM measurement}
Another practical challenge in our experimental implementation is related to the single-setting nature of the measurement stage. Indeed, our setup cannot directly perform a projective measurement in the OAM basis $\{\ket{m}\}_{m=-2}^2$.
Instead, we sequentially project onto each OAM state by using an SLM to transform that state into the fundamental TEM$_{00}$ mode, which is then coupled into a single-mode fiber and detected with an APD. 
Although formally the resulting statistics differs from that given by a direct projective measurement, we find that in practice this difference is negligible. 
We can track this down to the fact that with sufficiently high statistics, a multinomial distribution is well approximated by a set of Poisson-distributed detection events, rendering the sequential measurement effectively equivalent to a single projective measurement.

\subsection{Photon source}
\label{sec:photon_source}

\begin{figure*}[tb]
    \centering
    \includegraphics[width=0.8\linewidth]{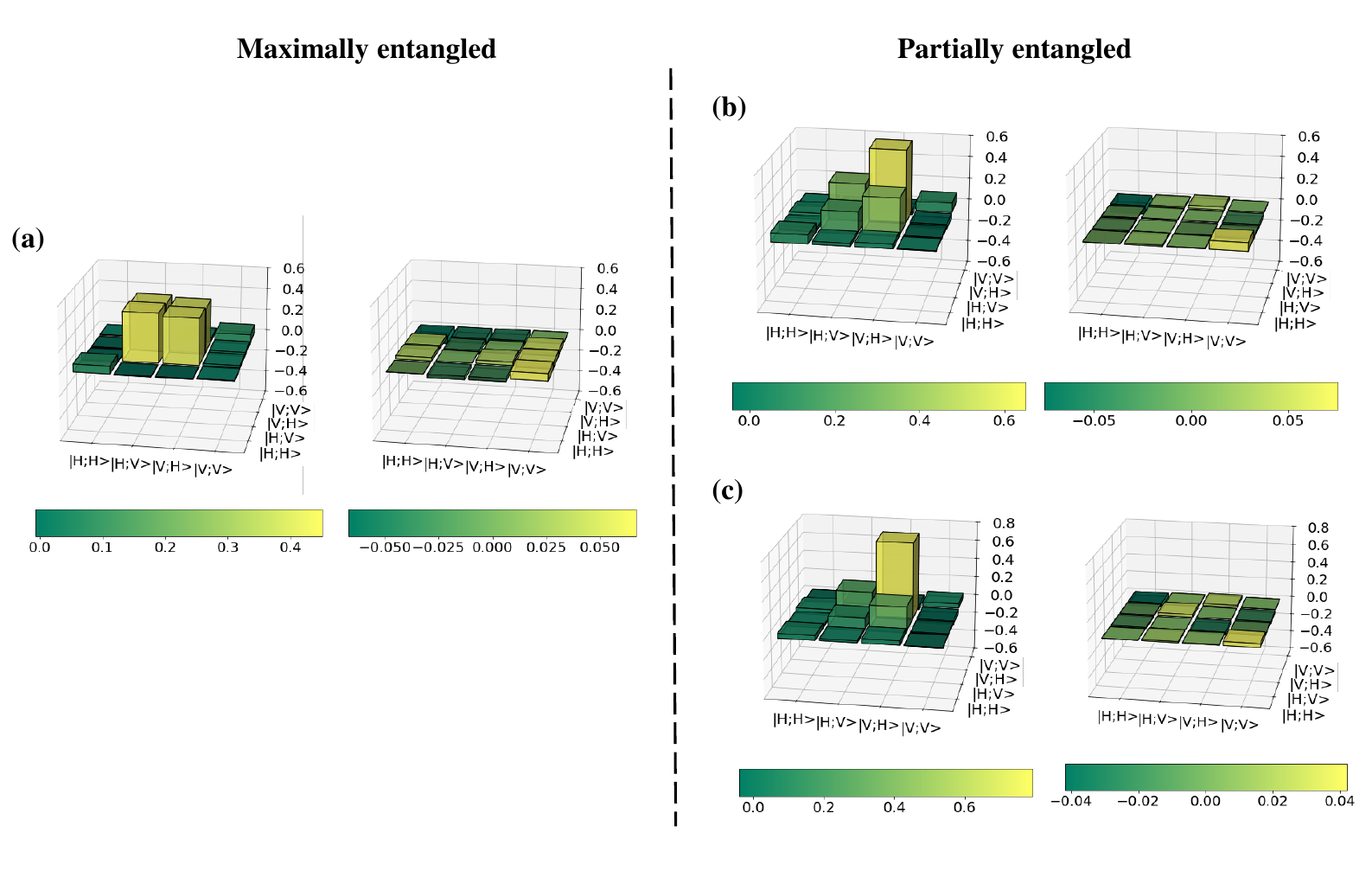}
    \caption{\textbf{Input states density matrix.} Experimentally reconstructed real (left) and imaginary (right) parts of the density matrix associated with the states generated by the SPDC source: (a) entangled $\ket{\Psi_R^+}$, (b) partially entangled $\ket{\Psi_R^+}_{p_1}$ and (c) partially entangled $\ket{\Psi_R^+}_{p_2}$.}
    \label{fig:Input_states}
\end{figure*}


\parTitle{Photon-pair generation}
The experimental setup used for generating input states is illustrated in Fig. 2-(a) of the Main Text.
Photon pairs at 808 nm are produced via spontaneous parametric down-conversion (SPDC) in a periodically poled potassium titanyl phosphate (ppKTP) nonlinear crystal, pumped by a continuous-wave laser at 404 nm and placed within a Sagnac interferometer.
The poling of the crystal is engineered to have a collinear emission with a type II phase-matching condition, enabling the process $\ket{H}_p \rightarrow \ket{H}_i\ket{V}_s$ where the subscripts denote the pump, idler, and signal photons.
By adjusting the pump polarization with a half-wave plate (HWP), various two-photon states can be generated, including both separable and entangled states.
We will refer to the states generated by the source as \textit{reference states}, as the ``input states'' that are fed into the model are obtained by applying local unitary operations to these reference states.
Changing the reference states thus allows to switch between generating separable or entangled states.
Separable states are obtained by pumping the crystal with a horizontally polarized beam, which generates photons in only one arm of the interferometer. 
As reference separable states we used $\ket*{\Psi_R^{\rm sep, 1}} \equiv\ket{H,V}$ and $\ket*{\Psi_R^{\rm sep, 2}} \equiv\ket{V,V}$, where the latter is obtain from the first by using an additional HWP before entering the QW setup.
Conversely, maximally entangled states are produced by using a diagonally polarized pump beam, resulting in a superposition of clockwise and counterclockwise emissions.
The maximally entangled states generated by our source have the form $\ket{\Psi^{+}_R} = \frac{1}{\sqrt{2}} ( \vert H \rangle_1 \vert V \rangle_2 + \vert V \rangle_1 \vert H \rangle_2 )$, where the subscripts indicate the output ports of the interferometers.
We furthermore investigated the generation and the model performance when the pump polarization is set in such a way as to generate \textit{partially} entangled reference states. In these cases, we used the states $\ket{\Psi_R^+}_{p_1} = 1/\sqrt{4}\ket{H}_1\ket{V}_2+\sqrt{3/4}\ket{V}_1\ket{H}_2$ and $\ket{\Psi_R^+}_{p_2} = 1/\sqrt{5}\ket{H}_1\ket{V}_2+\sqrt{4/5}\ket{V}_1\ket{H}_2$. The results of this analysis are reported in~\cref{sec:exp_results}.

\parTitle{Quality assessment of generated states}
To assess the quality of the states generated by the source we performed full quantum state tomography on them. In particular, we measured the expectation values of the combination of the three Pauli operators $\{\sigma_x, \sigma_y, \sigma_z\}$ and the identity $\mathbb{1}$ on the two-photon states, and used a maximum-likelihood approach to retrieve their density matrix. The real and imaginary parts of the reconstructed density matrices are shown in~\cref{fig:Input_states}. Moreover, we also verified the violation of the CHSH inequality to asses the entangled nature of the produced photon pairs.
The values obtained for the fidelities and the Bell parameter $S$ are collected in~\cref{tab:fid_Bell} for all the states under analysis.

\begin{table}
\centering
    \begin{tabular}{ c | c| c} 
    \hline
    State & $\mathcal{F}$ & $\mathcal{S}$ \\
    \hline
    $\ket{\Psi_R^+}$ & $0.923 \pm 0.002$ & $2.685 \pm 0.013$ \\ 
    \hline
    $\ket{\Psi_R^+}_{p_1}$& $0.905 \pm 0.002$ & $2.432 \pm 0.013$ \\ 
    \hline
    $\ket{\Psi_R^+}_{p_2}$& $0.927 \pm 0.001$ & $2.301 \pm 0.012$ \\ 
    \hline
    \end{tabular}
    \caption{\textbf{Quality of reference states.} We report here the fidelity $\mathcal{F}$, obtained from the experimentally reconstructed density matrix, and the Bell parameters $\mathcal{S}$ of the CHSH inequality violation for all the states under analysis. The errors are computed by assuming a Poissonian statistics on the experimentally measured counts.}
    \label{tab:fid_Bell}
\end{table}

\parTitle{Estimated experimental statistics}
The photon-pair source is pumped with a continuous-wave laser having a power $P \approx 8\,\rm{mW}$, and yielding a coincidence rate of $\rm{cc}_s\approx 20\,\mathrm{kHz}$ at the output of the Sagnac interferometer.
Using a coherent $808 \,\rm{nm}$ laser we measure the transmission efficiencies of all the elements present in the experimental setup. In particular, for the q-plates and the whole QW we obtained efficiencies $\eta_{\rm QP} \approx 0.80$ and $\eta_{\rm QW} \approx 0.56$, respectively.
At the QW output, we use a HWP, a QWP, and a PBS to project the polarization degree of freedom, which introduces an average signal loss of $\eta_{\rm proj} \approx 0.5$ per photon.
We then perform projective measurements in the OAM space by using a combination of an SLM and single-mode fibers coupling.
For these we measured average efficiencies of $\eta_{\rm SLM} \approx 0.78$ and $\eta_{\rm SMF} \approx 0.4$, respectively.
Combining all these factors, the estimated coincidence rate for each of the 25 output events is
\begin{equation}
    \mathrm{cc}_{\rm teo} = \eta_{\rm QW}^2 \eta_{\rm proj}^2 \eta_{\rm SLM}^2 \eta_{\rm SMF}^2 \mathrm{cc}_s \approx 6.1\,\mathrm{Hz}.
\end{equation}
By comparison, for the \sfE1 experimental configuration we observed a rate of $\approx 2.9\,\mathrm{Hz}$. In this experiment, we set the acquisition time to 8 seconds per projection, then, repeating each measurement 2 times to increase the statistical significance.
The overall average counts per state is then found to be $1117 \pm 76$ cc/state, summing the observed counts over all 25 outcomes.

\subsection{Experimental configurations}
\label{sec:reservoir_configurations}

We report here a summary of the experimental configurations used to acquire the reported data.
We considered several scenarios to test how performances are affected by different model parameters, choice of training states, and collected statistics.

\parTitle{Experimental settings}
We present data relative to 6 distinct experimental datasets, labeled \sfE1, \sfE2, \sfE3, \sfE4, \sfE5, \sfE6.
These correspond to different reservoir and training configurations.
We used three reservoir configurations, labeled \sfR1, \sfR2, and \sfR3.
\resOptimal is the reservoir obtained setting the quantum walk waveplates angles in order to minimize the trace of the inverse frame superoperator corresponding to the effective POVM.
This is a procedure previously outlined in [52] that provides the reservoir configuration corresponding to the minimal MSE averaged over all possible target observables.
\resFakeOptimal is a reservoir configuration obtained from \resOptimal by swapping the angles of the QWPs implementing the coin operator in the second step of the QW evolution, and \resRandom is yet another reservoir configuration obtained setting the waveplates angles totally at random.
A breakdown of the characteristics of each dataset presented is provided in~\cref{table:experiments_summary}.
The exact values used for the reservoir configurations, as well as the experimental data used to train and test the QELM in each configuration, and the code used to generate the reported data, is available at~\url{https://github.com/salvatore-lorenzo/arXiv-2502.18361-QELM\_WITNESS}.

\parTitle{Input states}
The input states used to train the model were generated by applying local unitary operations --- realized with a HWP-QWP pair on each QW --- to the reference states.
Each such single-qubit unitary has thus the form $U(\phi,\theta)=\mathrm{QWP}(\phi)\mathrm{HWP}(\theta)$, for some pair of angles $\phi,\theta\in\mathbb{R}$.
Tuning the angles in each of these four waveplates we can generate a large variety of possible input states.
In all cases, we generated the input states sampling uniformly random angles for each of the waveplates.
For our results, we either used the same angles on both sides, or we sampled the random angles independently on the two arms.
In the first case, the input states thus have the form $(U(\phi,\theta)\otimes U(\phi,\theta))\ket*{\Psi_R}$ with $\ket*{\Psi_R}$ the reference state.
This is the scenario considered in the datasets \sfE1, \sfE2, \sfE3, \sfE5, \sfE6.
On the other hand, in \sfE4 we analyzed the performances in the case where input states take the form $(U(\phi_1,\theta_1)\otimes U(\phi_2,\theta_2))\ket*{\Psi_R}$, with independently sampled $\phi_1,\theta_1,\phi_2,\theta_2$.

\parTitle{Singular values and model trainability}
A practical way to assess the structure of a set of states from the perspective of model trainability is to examine the dimension of the space it spans.
This can be done by inspecting the singular values of the matrices whose rows are the vectorized density matrices.
Focusing on the cases where training and test states are comprised entirely of separable and entangled states, respectively, we thus define the matrices $M^{\rm sep}$, $M^{\rm ent}$, and $M^{\rm all}$ to contain separable, entangled, and all states, respectively.
More precisely, $M^{\rm sep}$ is the matrix whose \(k\)-th row is the vectorization of the $k$-th separable state.
The number of nonzero singular values of $M^{\rm sep}$, which also equals \(\on{rank}(M^{\rm sep})\), is then the dimension of the real linear span of \(\{\rho^{\rm tr}_k\}_{k=1}^{N_{\rm tr}}\), when the training is comprised entirely of separable states.
We similarly define \(M^{\rm ent}\) and \(M^{\rm all}\) for the entangled states alone and for all states, respectively.
This yields the three numbers \(\on{rank}(M^{\rm sep})\), \(\on{rank}(M^{\rm ent})\), and \(\on{rank}(M^{\rm all})\), reported in~\cref{table:experiments_summary} for our various experimental configurations. 
These ranks reveal whether successful QELM training can be expected from a fundamental informational standpoint.
In particular, \(\on{rank}(M^{\rm sep}) = 16\) means that the training separable states span the entire 16-dimensional space of density matrices in our setup, enabling the QELM to learn any feature of any state. 
Conversely, if \(\on{rank}(M^{\rm sep}) < \on{rank}(M^{\rm all})\), which happens for \sfE2, \sfE3, \sfE5, \sfE6, certain features of the test states lie outside the linear span of separable states, requiring to add some entangled states in the training.
Ideally, \(\on{rank}(M^{\rm sep}) = \on{rank}(M^{\rm all})\), as is true for the \sfE1 and \sfE4. 
On the other hand, higher \(\on{rank}(M^{\rm sep})\) corresponds to a higher resource requirement in terms of training statistics used per state.
The best configuration choice thus ultimately depends on balancing these resource demands against the actual application one has in mind for the device. 

\parTitle{Polarization projection}
The fixed polarization state $\ket{\eta}=\cos(\theta_p)\ket{H}+e^{i \phi_p}\sin(\theta_p)\ket{V}$, on which the QW output states are projected before measuring their OAM distribution, is given by the same minimization on the frame superoperator reported above [52]. In particular, using a HWP, a QWP and a PBS to perform the polarization projection, we set the waveplates angles $\{\theta_{\rm proj},\phi_{\rm proj}\}$ by solving the relation:
\begin{equation}
\on{QWP}(\theta_{\rm proj})
\on{HWP}(\phi_{\rm proj})\ket{\eta}=\ket{H}
\end{equation}
Therefore, using the PBS to allow only the transmission of H-polarized photons, we actually perform a polarization projection on the wanted $\ket{\eta}$ state.

\section{Additional experimental results}
\label{sec:exp_results}

\begin{figure*}[tb]
    \centering \includegraphics[width=0.9\linewidth]{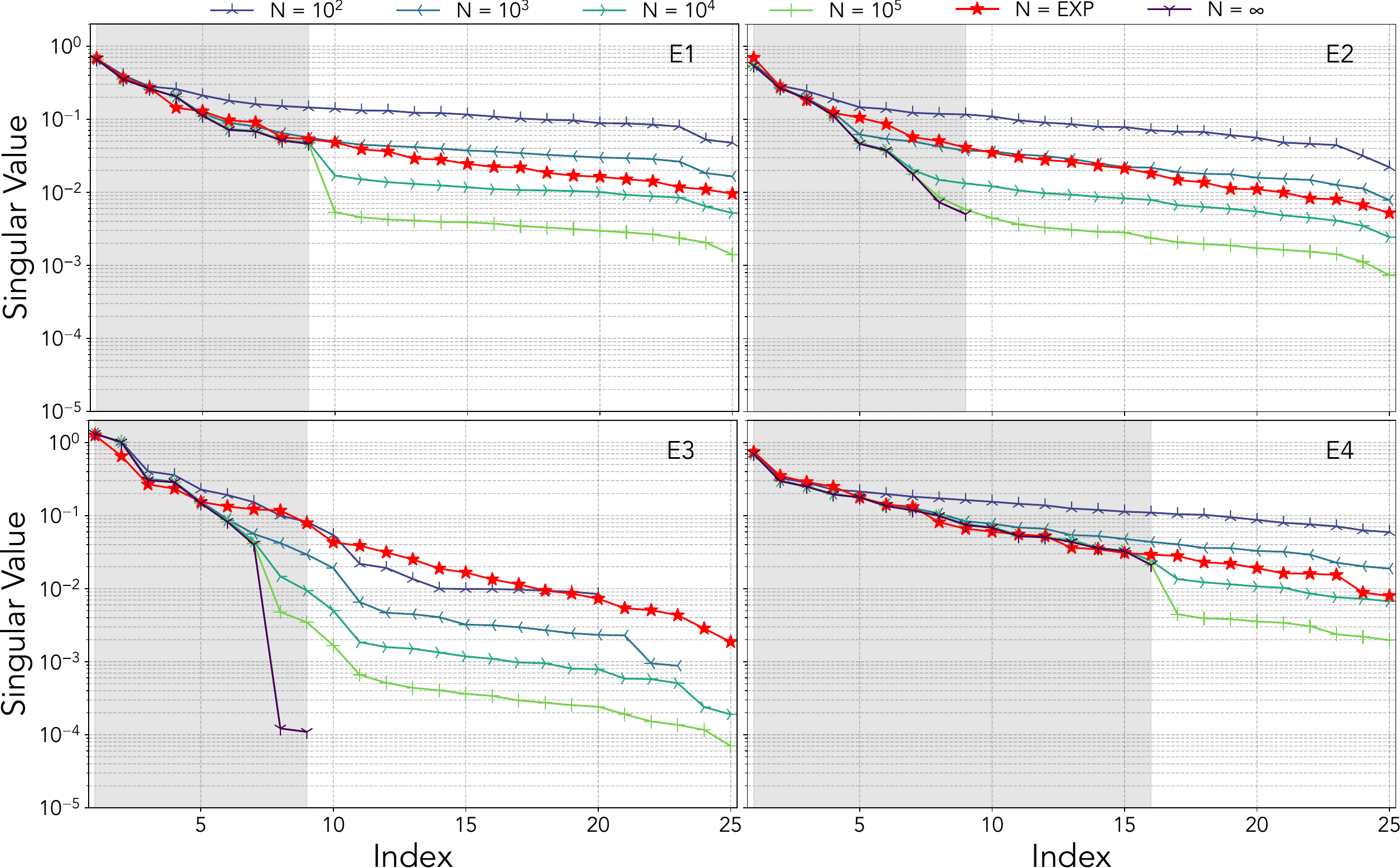}
    \caption{
    \textbf{Theoretical vs experimental singular values.}
    Singular values of the probability matrix $\langle\bstildemu,\bs\rho\rangle$ computed from the separable states in each of the four configurations \sfE1, \sfE2, \sfE3, \sfE4.
    For each configuration, we plot the singular values of the matrix obtained by sampling from $\langle\bstildemu,\bs\rho\rangle$ with statistics $N$, along with the $N=\infty$ (which corresponds to exact probabilities), and the singular values of the corresponding experimental counts matrix.
    In the experimental case, the matrix is normalized by the estimated number of states before the polarization projection, calculated as $N=4 \overline N$, where $\overline N$ is the averaged observed counts over all separable states.
    The shaded region highlights the number of nonzero singular values for the exact probability matrix, and thus corresponds to 
    \(\on{rank}(M^{\rm sep})\) 
    as reported in~\cref{table:experiments_summary}.
    As \(N\) increases, the sampled singular values approach those of the \(N = \infty\) curve.
    Configurations \sfE1 and \sfE4 use the \resOptimal reservoir (which is optimized for information recovery), and correspondingly exhibit higher singular values, whereas \sfE2 and \sfE3 use less optimal reservoirs, and thus correspondingly smaller singular values.
    In \sfE3, for instance, even \(N = 10^5\) is insufficient to reproduce the smallest two singular values, implying the existence of two observables which to be recovered require an even larger amount of training statistics.
    The singular values from the experimental counts matrix roughly align with theoretical curves at \(N \sim 10^3\), matching the typical count rates observed in the experiment.
    }
    \label{fig:singular_values}
\end{figure*}

\parTitle{Singular values of counts matrices}
A useful way to assess how readily information can be extracted from experimental data, given a choice of training states and reservoir, is to examine the spectrum of singular values of the matrix
\(\langle \bstildemu, \rho^{\rm tr} \rangle\),
and its sampled counterpart obtained for different training statistics \(N\).
This analysis is presented in~\cref{fig:singular_values}, where we focus on the case of exclusively separable training states.
When using exact probabilities and assuming an informationally-complete effective POVM, the number of nonzero singular values equals 
\(\on{rank}(M^{\rm tr})\).
Although this rank reveals the theoretical maximum amount of information contained in the measurement data, the magnitudes of the singular values themselves are equally important. 
In particular, small singular values correspond to observables that require large training statistics for accurate reconstruction. 
This viewpoint aligns with [47], where the condition number of the same matrix is used to quantify the ill-conditioning of the QELM training problem.
Note that 
\(\langle \bstildemu, \rho^{\rm tr} \rangle\) 
can have at most 16 nonzero singular values, corresponding to the dimension of the space of input density matrices.
However, when sampling with finite $N$, we get more than 16 nonzero singular values due to sampling noise.
Consequently, any singular values of similar order to those beyond the 16th effectively correspond to observables that cannot be accurately estimated from the available measurement data, since the associated data is indistinguishable from the statistical noise.

\begin{figure*}[tb]
    \centering \includegraphics[width=0.8\linewidth]{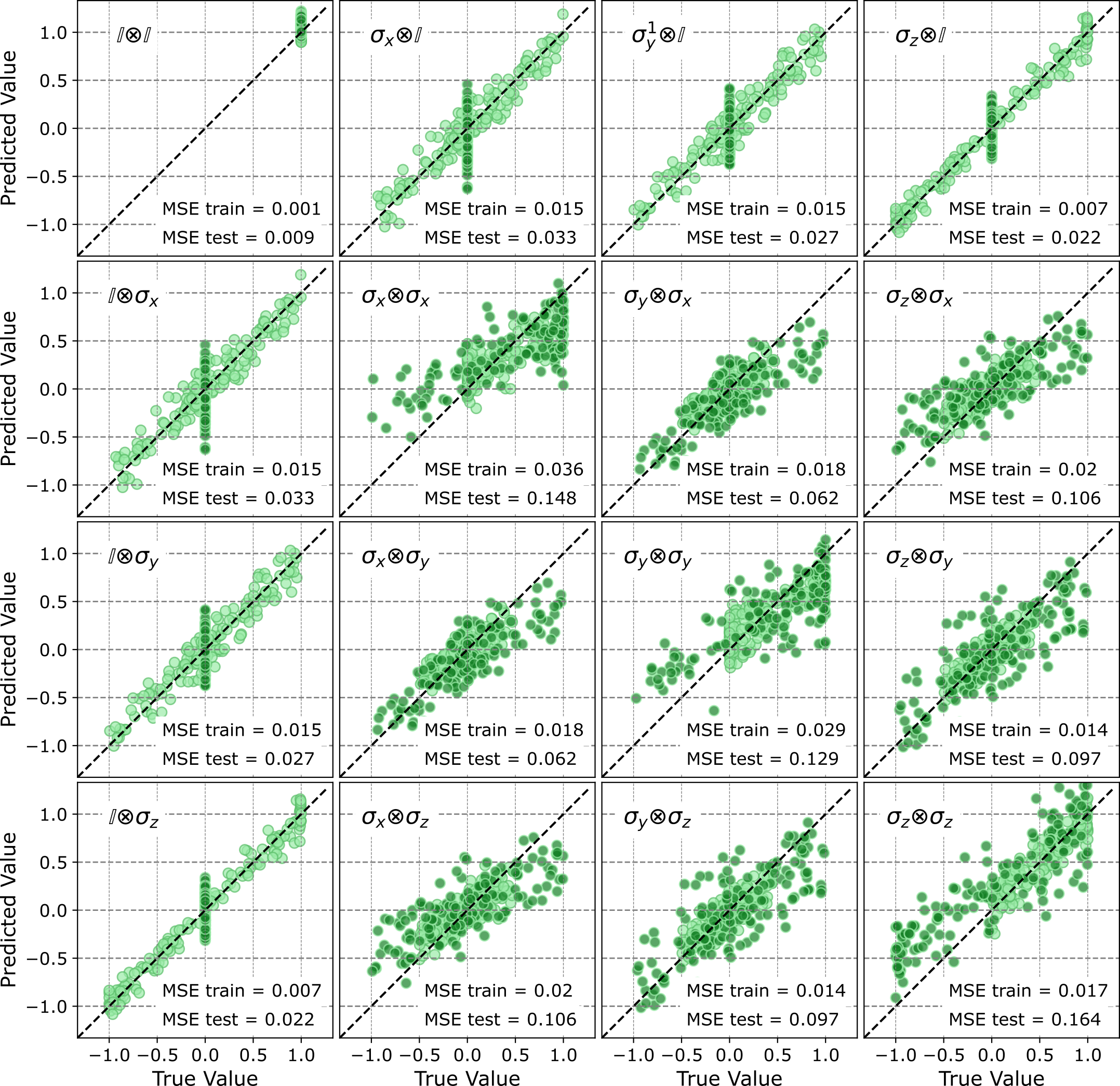}
    \caption{
    \textbf{Reconstruction MSEs of Pauli observables.}
    Estimation performance in \sfE1 for target observables that are products of local Pauli matrices, i.e., \(\sigma_j \otimes \sigma_k\) with \(j, k \in \{0,1,2,3\}\) and \(\sigma_0 \equiv I\).
    In all cases, the training set consists entirely of separable states, while the test set consists entirely of maximally entangled states.  
    Each plot displays both the training predictions (light-green dots) and the test predictions (dark-green dots).  
    For local Pauli observables of the form \(\sigma_i \otimes I\) or \(I \otimes \sigma_i\), \(i \in \{1,2,3\}\), the true expectation values are zero, resulting in the observed vertical clustering of predicted values for the maximally entangled states.
    Within each plot, we report the average MSEs for training states and for test states.
    These results clearly show that the estimation MSE depends on the choice of target observable, in agreement with theoretical expectations.
    In the limit of very large training and test sample sizes, we expect all points in these scatter plots to align along the main diagonal.
    }
    \label{fig:E1_paulis}
\end{figure*}

\begin{figure*}[tb]
    \centering \includegraphics[width=0.9\linewidth]{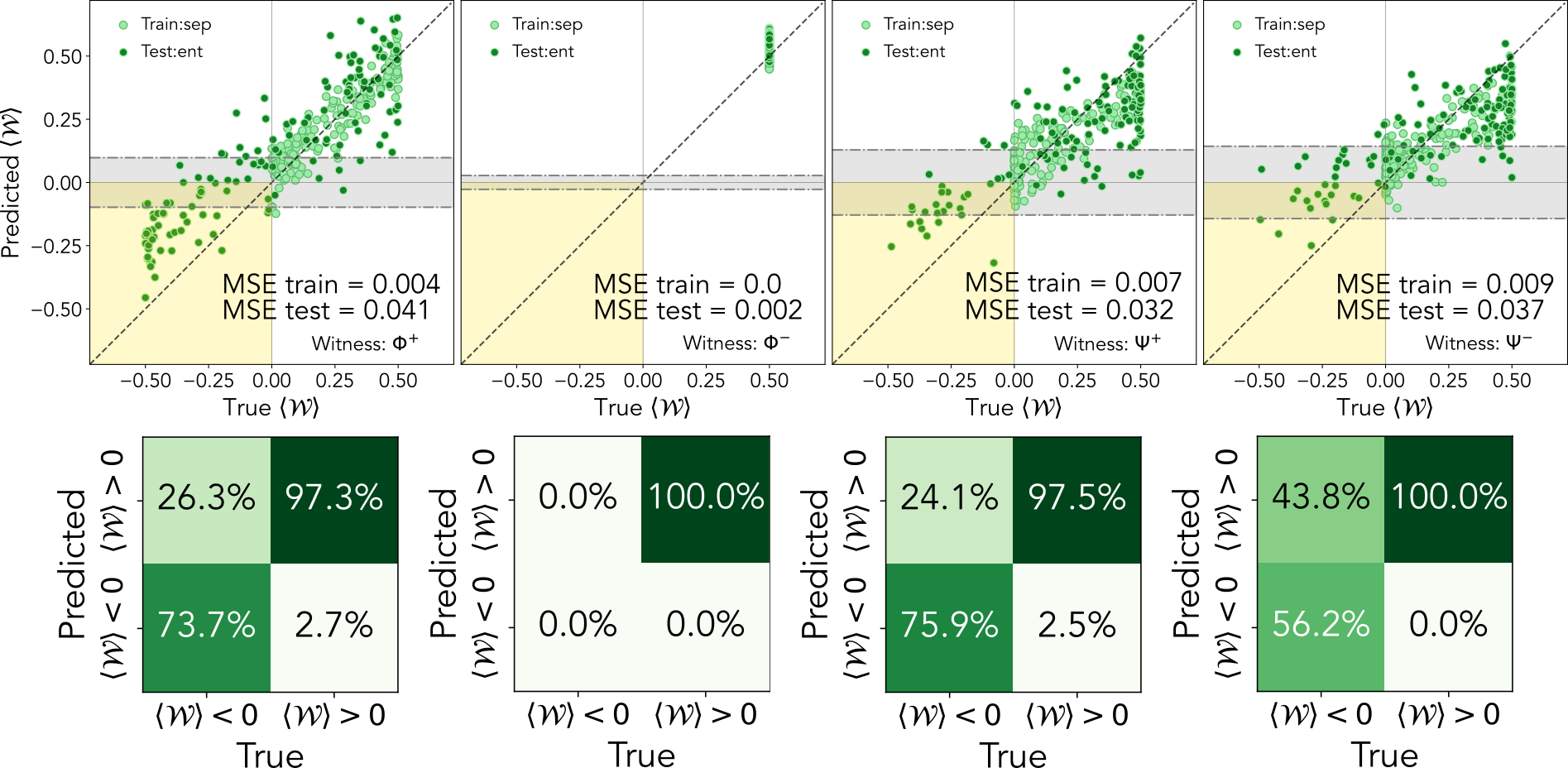}
    \caption{
    \textbf{Reconstruction MSEs for entanglement witnesses.}
    Estimation performance in \sfE1 for the four Bell witnesses $\calW_i$.
    As before, the training contains only separable states, while the test contains only maximally entangled states.  
    Each plot shows both the training predictions (light-green dots) and the test predictions (dark-green dots).
    The results for the $\Phi^+$ witness are the ones also reported in the main text.
    The trivial results for the $\Phi^-$ witness are due to the particular symmetry of states generated in the \sfE1 configuration, which all give the same value for $\langle\mathcal W_{\Phi^-}\rangle$.
    Within each plot, we report the mean squared error (MSE) averaged over all training states and all test states.
    Furthermore, the shaded gray area has width corresponding to the average MSE estimated from the training dataset.
    Points with predicted $\langle\calW\rangle$ below this region can be more reliably considered as certifiably entangled.
    }
    \label{fig:E1_witnesses}
\end{figure*}

\begin{table*}
\centering 
\begin{tabular}{c|c|c|c|c|c}
\hline
\textbf{Label} & \textbf{Reservoir type} & \textbf{Reference entangled states} & \textbf{Reference separable states} & \textbf{Generation angles} & \textbf{States ranks} \\ \hline
\sfE1 & \resOptimal & $\ket*{\psi^+_R}$ & $\ket*{VV}$ & Equal & 9-9-9 \\ \hline
\sfE2 & \resFakeOptimal & $\ket*{\psi^+_R}$, $\ket*{\psi^+_R}_{p1}$ & $\ket*{VH}$ & Equal & 9-6-10 \\ \hline
\sfE3 & \resRandom & $\ket*{\psi^+_R}$ & $\ket*{VH}$ & Equal & 9-6-10\\ \hline
\sfE4 & \resOptimal & $\ket*{\psi^+_R}$ & $\ket*{VH}$ & Different & 16-10-16 \\ \hline
\sfE5 & \resFakeOptimal & $\ket*{\psi^+_R}$ & $\ket*{VH}$ & Equal & 9-6-10 \\ \hline
\sfE6 & \resFakeOptimal  & $\ket*{\psi^+_R}$, $\ket*{\psi^+_R}_{p2}$ & $\ket*{VH}$ & Equal &9-6-10
\\
\hline
\end{tabular}
\caption{
    \textbf{Summary of experimental configurations}
    For each configuration $\mathsf Ei$ we report which of the three reservoir configurations was used; the reference entangled and separable states; whether the preparation angles used to generate input states were equal or different for the two QWs; under ``states ranks'', the ranks of \(M^{\rm sep}\), \(M^{\rm ent}\), and \(M^{\rm all}\), corresponding to the dimension of the span of separable and entangled states, and of all states.
}
\label{table:experiments_summary}
\end{table*}


\parTitle{Additional data for \sfE1}
We report in~\cref{fig:E1_paulis,fig:E1_witnesses} the estimation performances using Pauli matrices as target observables, as well as for the four entanglement witnesses built with the Bell states, in \sfE1.
In particular,~\cref{fig:E1_paulis} provides a more complete picture of how our apparatus maps input information into the measured degrees of freedom.
We observe that local observables are predictably easier to reconstruct when using separable states. The vertical lines corresponding to the test entangled state are due to testing states being maximally entangled, meaning they always have zero expectation value on local Pauli observables.

\begin{figure*}[tb]
    \centering \includegraphics[width=1\linewidth]{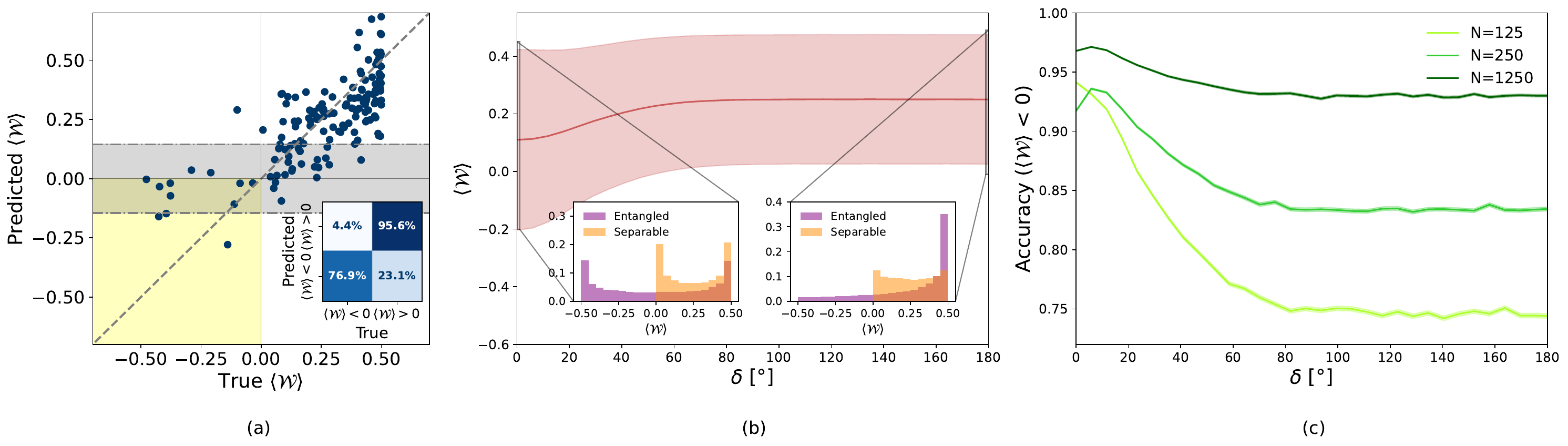}
    \caption{
    \textbf{Effect of waveplates angles mismatch.}
    Estimation performances to reconstruct $\mathcal W_{\Phi^+}$ as we move from the same angle condition, by adding a mismatch extracted from a normal distribution $\mathcal{N}(0,\delta^2)$ in the input stage HWP and QWP of one of the two quantum walks (see Fig. 2 of the main text).
    (a) Estimation performance in \sfE4. We collected 300 quantum states, dividing them into training and test sets of 150 elements. The plot shows the retrieved witness values and the confusion matrix over the test set. This dataset achieves values of the MSE equal to 0.021 and 0.022, respectively for the training and test set, where the square root of the former is also reported as a gray shaded area in the plot. 
    (b) The plot shows the theoretical witness expected values for the states in the dataset as a function of the angle mismatch $\delta$, where the red solid line and shaded area respectively represent their mean and standard deviation. The purple (orange) histogram corresponds to the witness estimations over the entangled (separable) states, and it showcases how the overall distribution moves towards positive values when $\delta$ increases, reducing in this way the negative instances in the dataset.
    (c) Theoretically simulated estimation performance in retrieving negative values over the test set. The plot reports the mean accuracy over 5000 angles mismatch instances (randomly selected from $\mathcal{N}(0,\delta^2)$), for three different statistics conditions (125, 250, 1250 signals). The results highlight that the estimation accuracy tends to decrease when $\delta$ increases, with a contribution that is less relevant for higher statistics. The error band associated with each of the three curves is calculated as the standard deviation of the mean, which is too small to be evident in the plot.
    }
    \label{fig:QELM_diff_angles}
\end{figure*}

\parTitle{Training with asymmetric input states}
We next investigated estimation performances in \sfE4 (cf. \cref{table:experiments_summary}).
In this case we used a input state preparation scheme where the waveplate angles used to prepare the input states of the two QWs are chosen independently from each other.
This naturally increases both the complexity of the estimation task and the required experimental statistics. 
We once again evaluate the estimation accuracy for various entanglement witnesses and Pauli observables. 
In~\cref{fig:QELM_diff_angles}-(a) we show that \(\mathcal{W}_{\Phi^+}\) is still estimated effectively in this scenario, though the performance is somewhat lower than in \sfE1. 
This reduction arises from the higher training-statistics requirement imposed by a larger training set, rather than from any fundamental limitation of the method. 
\Cref{fig:QELM_diff_angles}-(b) reports the exact expectation values \(\langle\mathcal{W}_{\Phi^+}\rangle\) for random states, where the waveplate angles used to prepare the input states entering the two QWs differ by an amount drawn from a normal distribution \(\mathcal{N}(0,\,\delta^2)\). 
Under these conditions, fewer of the generated states exhibit negative \(\langle\mathcal{W}_{\Phi^+}\rangle\). 
Overall,~\cref{fig:QELM_diff_angles} illustrates how the accuracy of entanglement witnessing declines when states are generated at varying angles and the available statistics are limited. 
Nevertheless, the same performance attained in \sfE1 can be recovered with a larger training-data budget, demonstrating that the drop in accuracy is not intrinsic but rather a consequence of increased training demands.

\begin{figure*}[tb]
    \centering \includegraphics[width=0.9\linewidth]{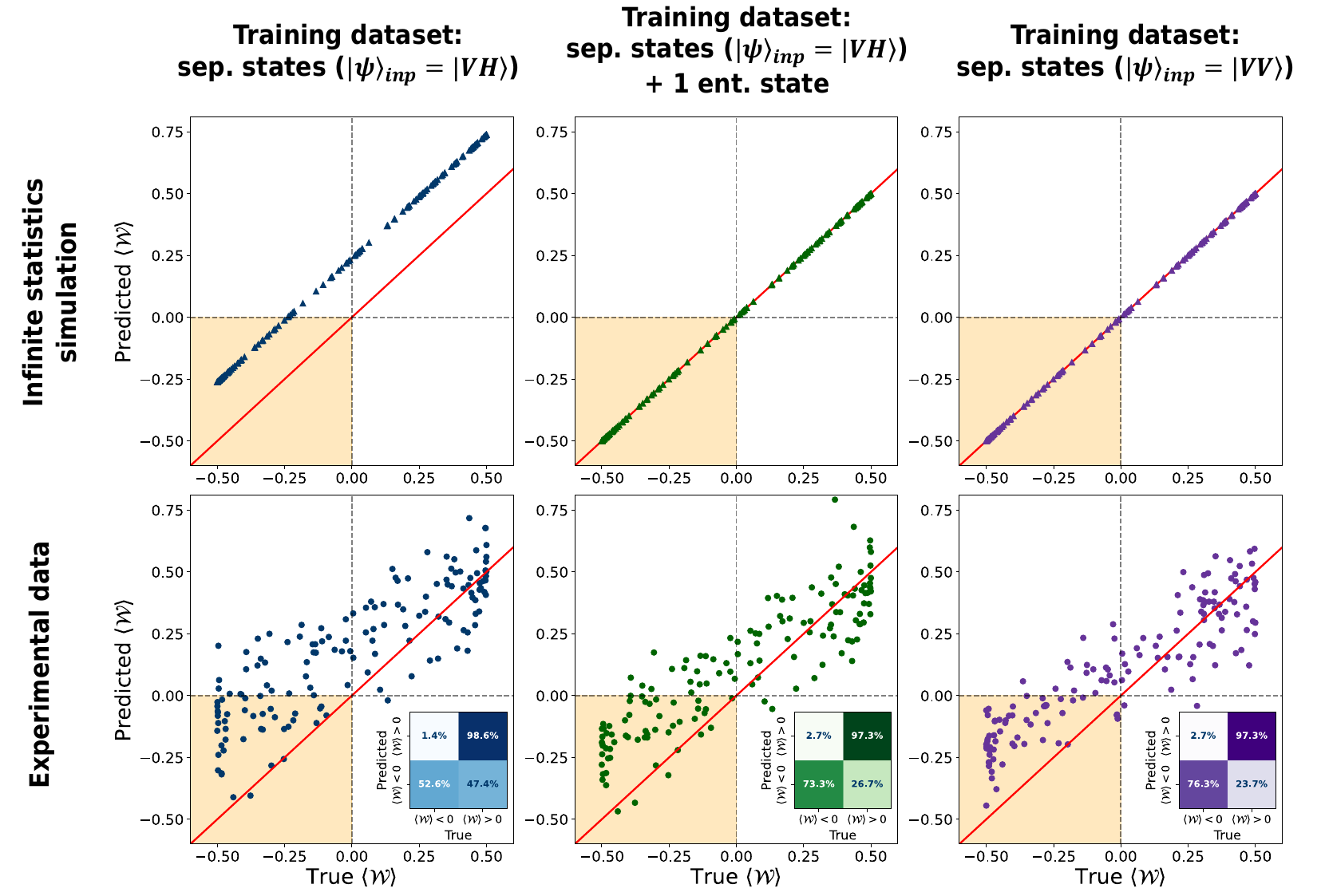}
    \caption{\textbf{Predictions with different separable input states.}
    Estimation performances to reconstruct $\mathcal W_{\Phi^+}$ training with separable states and testing with entangled states. The upper row reports the simulated case with infinite training statistics, while the lower displays the experimental one.
    The experimental data in the first two plots comes from \sfE5, while the third from \sfE1. The first column reveals how, also in the infinite statistic case scenario, using  $\ket{HV}$ as input separable state doesn't enable us to retrieve the correct values of the witness, resulting in a constant shift in the estimation. This can be solved by adding just one entangled state to the training set (middle column), as evident also from the accuracy performance in the experimental regime.
    The third column shows that using $\ket{VV}$ as separable reference state allows to correctly estimate $\mathcal W_{\Phi^+}$.}
    \label{fig:QELM_diff_sep}
\end{figure*}

\parTitle{Training with different reference separable states}
We also investigated how the estimation performance changes when a different reference separable state is used prior to the state preparation stage.
A characteristic of the QW dynamics we employ is that, when using the reference state $\ket{VH}$, as done in 
\sfE2-\sfE6, only a strict subspace of states is generated when the same angles are used on both preparation branches, due to the specific symmetries of the dynamics.
In this scenario, the input states have the form $(U \otimes U)\ket{VH}$ and $(U \otimes U)\ket*{\Psi_R^+}$ for separable and entangled states, respectively, where $U$ is a random single-qubit unitary.
This can be verified by studying the ranks of $M^{\rm sep}$ and $M^{\rm all}$, which are also reported in~\cref{table:experiments_summary}.
When the reference separable state is $\ket{VH}$, we have $\on{rank}(M^{\rm all}) > \on{rank}(M^{\rm sep})$, indicating that some test entangled states span directions the QELM never had a chance to learn about during training.
This effect is showcased in~\cref{fig:QELM_diff_sep}, where we observe that even in the theoretical simulation there is a shift between the true and predicted expectation values of the target witness $\calW_{\Phi^+}$.
This constant bias arises because all entangled states in the test share a fixed expectation value on the single observable present in the test but missing in the training dataset.
In fact, adding even just one single entangled state to the training dataset is sufficient to characterize that missing dimension, and thus fix the bias.
On the other hand, this issue does not arise when the reference separable state is $\ket{VV}$, owing to the different symmetry properties of the reservoir in this case.
These findings represent further evidence that the way training states are generated can be crucial from an estimation perspective, particularly when the underlying reservoir dynamics present specific symmetries and do not uniformly spread information at random into the output degrees of freedom.

\begin{figure*}[tbh]
    \centering \includegraphics[width=\linewidth]{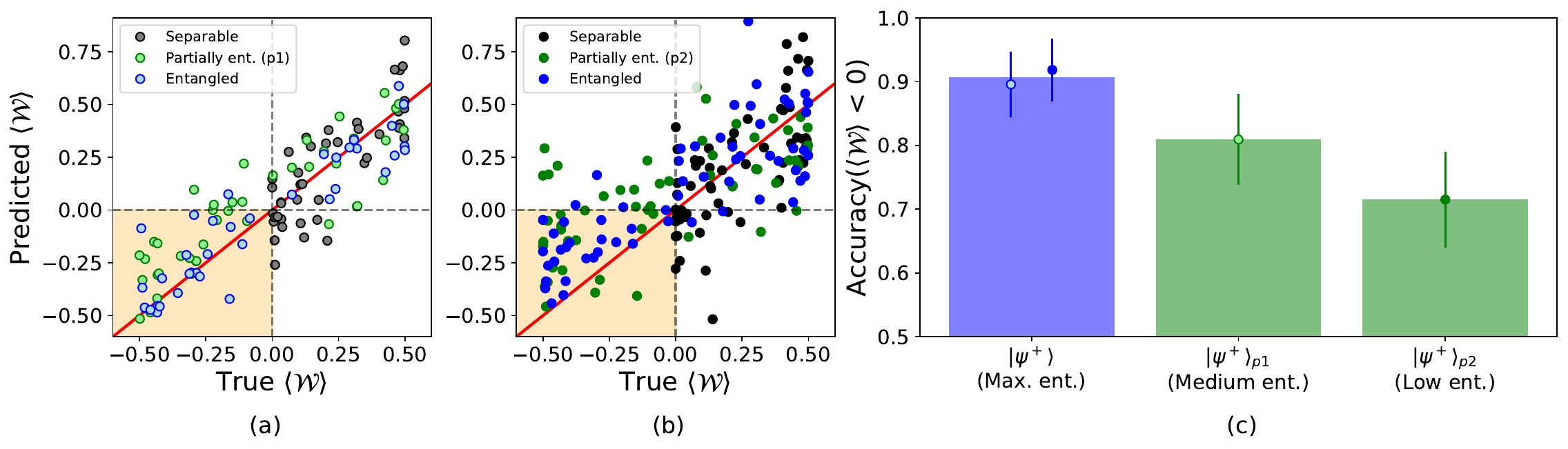}
    \caption{
    \textbf{Estimation performance with partially entangled reference states.}
    Panels \textbf{(a)} and \textbf{(b)} show the predicted expectation values of $\mathcal W_{\Phi^+}$ when using as reference entangled states the partially entangled $\ket*{\Psi^+}_{p_1}$ and $\ket*{\Psi^+}_{p_2}$, as defined in~\cref{sec:photon_source}.
    More specifically, we intentionally mislabel the partially entangled states during training, to assess the model's robustness to mislabeling errors.
    Panel \textbf{(c)} reports the accuracy in detecting states with $\langle\mathcal W_{\Phi^+}\rangle<0$ for the three entanglement cases.
    The bar heights show the average accuracy over $10^4$ random splits of the training and test sets, with the error bars representing the corresponding standard deviations.
    Green bars correspond to partially entangled states, while the blue bar represents maximally entangled states.
    Within each color, empty dots refer to the $p_1$ dataset and filled dots to the $p_2$ dataset.
    The heights of the green bars reflect the average accuracy specific to $\ket*{\Psi^+}_{p_1}$ and $\ket*{\Psi^+}_{p_2}$, whereas the blue bar's height is the mean accuracy across both datasets for the maximally entangled states.
    }
    \label{fig:QELM_parz_ent}
\end{figure*}

\parTitle{Training with partially entangled reference states}
As discussed in~\cref{sec:photon_source}, the Sagnac source allows to produce photon pairs with varying degrees of entanglement by adjusting the pump polarization.
We leveraged this to also investigate, in \sfE2 and \sfE6, the estimation performance when using partially entangled states, focusing in particular on the states $\ket*{\Psi^+}_{p_1}$ and $\ket*{\Psi^+}_{p_2}$ described above.
For each of these reference states, we collected a dataset of 249 states in the first case and 357 in the second, both equally divided among maximally entangled, partially entangled, and separable states. Both these datasets are generated considering the \resFakeOptimal reservoir configuration and using identical angles at the state preparation stage.
We denote these two datasets as $p_1$ and $p_2$, respectively, indicating which partially entangled state they contain.
To test the model's resilience to undetected experimental defects in entangled-photon generation, we emulate an imperfect source that randomly produces either maximally or partially entangled photon pairs, without our knowledge.
We use a training dataset comprising 50\% of the total dataset and intentionally mislabel the partially entangled states as maximally entangled.
The predicted expectation values of $\mathcal{W}_{\Phi^+}$ are shown in~\cref{fig:QELM_parz_ent}-(a,b) for the $p_1$ and $p_2$ datasets, respectively.
Moreover, we examine how varying degrees of entanglement in the states $\ket*{\Psi^+}$, $\ket*{\Psi^+}_{p_1}$, and $\ket*{\Psi^+}_{p_2}$, affect the model performance.
As reported in~\cref{fig:QELM_parz_ent}-(c), the accuracy remains high across all three classes of entangled states, even for those with $\expval{\mathcal{W}_{\Phi^+}}<0$.
These results indicate that robust estimation performance can still be maintained even when undetected imperfections reduce the amount of entanglement.

\parTitle{Additional data}
Additional experimental results, including the MSEs for all Pauli observables and Bell witnesses for all the reported experimental configurations, as well as the code used to generate all reported data, is available in the GitHub repository~\url{https://github.com/salvatore-lorenzo/arXiv-2502.18361-QELM\_WITNESS}.

\clearpage
\section{Benchmarks with alternative approach}
\label{sec:benchmark_alternative_approaches}

\parTitle{Description of shadow tomography}
To benchmark the estimation performances of our QELM-based approach it is useful to compare them with what we would have obtained with alternative methods relying on prior knowledge of the experimental apparatus. 
In particular, shadow tomography [5,8,9,12] is a general methodology for computing estimators of target observables, in any measurement scenario that can be described by a POVM applied to output states after some evolution. 
More specifically, shadow tomography provides the optimal unbiased estimator --- meaning that it produces the unbiased estimator with the lowest possible average variance. 
Shadow tomography thus serves as a natural benchmark for QELM-based estimation, since both approaches aim to enable efficient estimation of specific properties of input states without requiring a full tomographic reconstruction.

\parTitle{How does shadow tomography work}
The core difference between shadow tomography and QELM-based reconstruction lies in the way the estimator is constructed. 
In QELMs, we use a training dataset of states and the corresponding measurement outcomes. 
In contrast, shadow tomography --- like state-tomography techniques --- requires no training dataset but instead demands explicit and accurate knowledge of the entire measurement apparatus.
Formally, we can compute the shadow tomography estimator associated with any POVM \(\bstildemu\). 
In our case, \(\bstildemu\) represents the effective measurement describing everything that happens to the input states, up to the final measurement. 
Given this POVM, we define the \emph{frame superoperator} \(\mathcal{F}\) as
$
  \mathcal{F}(X) \;\equiv\; \sum_{b} \mathrm{Tr}\bigl(\tilde{\mu}_b\,X\bigr)\,\tilde{\mu}_b,
$
for any operator \(X\).
We then compute the \emph{dual measurement frame} elements
\(\tilde{\mu}_b^\star \equiv \mathcal{F}^{-1}(\tilde{\mu}_b)\), where \(\mathcal{F}^{-1}\) is the inverse of \(\mathcal{F}\), understood as a linear operator acting on the space of Hermitian operators. 
As shown in [9], \(\tilde{\mu}_b^\star\) can be interpreted as the optimal unbiased estimator for the state. 
Given any target observable \(\mathcal{O}\), the corresponding optimal unbiased estimator is then defined by
$
  \hat{o}(b) \;\equiv\; \mathrm{Tr}\bigl(\mathcal{O}\,\tilde{\mu}_b^\star\bigr).
$

\parTitle{How to apply shadow tomography in our case}
Concretely, estimating target observables via shadow tomography in our experiment would thus entail computing analytically the effective POVM $\bstildemu$, and use it to build $\tilde\mu_b^\star$ with the recipe outlined above.
The reservoir dynamic we employ, as previously discussed in~\cref{sec:reservoir_configurations}, has the form
$(I_{\rm OAM}\otimes \bra{\eta})V\ket\Psi$, with $\ket{\eta}$ the polarization state on which we project at the end, $\ket\Psi$ the input bipartite polarization state, and $V\equiv V_1\otimes V_2$ the isometry corresponding to the two QWs.
The corresponding effective POVM is given by
$\tilde\mu_b = V^\dagger(\ketbra b\otimes\ketbra{\eta}) V$,
where $\ket b$ is the OAM computational basis, and this is then used to compute $\tilde\mu_b^\star$.

\parTitle{Knowing counts vs frequencies}
We previously discussed in~\cref{sec:thedeparted} why and how QELMs can operate without explicit knowledge of the total input statistics \(N\), relying solely on raw detection counts. 
This issue is exacerbated for alternative methods such as shadow tomography or standard tomography, where there is no training dataset to help estimate $N$.
Consequently, one must in these cases rely on an experimental estimate of the overall transmission rate, which might bring up additional estimation errors.
To avoid an unfair comparison and artificially pessimistic outcomes for shadow tomography, we instead report the MSE as a function of the guessed \(N\) and then take its minimum value.
It is to be noted, however, that this minimum depends on the target observable, implying different values of $N$ for different observables --- an obviously unphysical scenario.
Hence, the values obtained should be viewed as lower bounds on what would be achieved if the true \(N\) were known. 
In other words, they represent a best-case scenario --- that is, a lower bound --- for the performance of linear-estimator-based methods that rely on measurement probabilities at the output.

\begin{figure}[tb]
    \centering
    \includegraphics[width=1\linewidth]{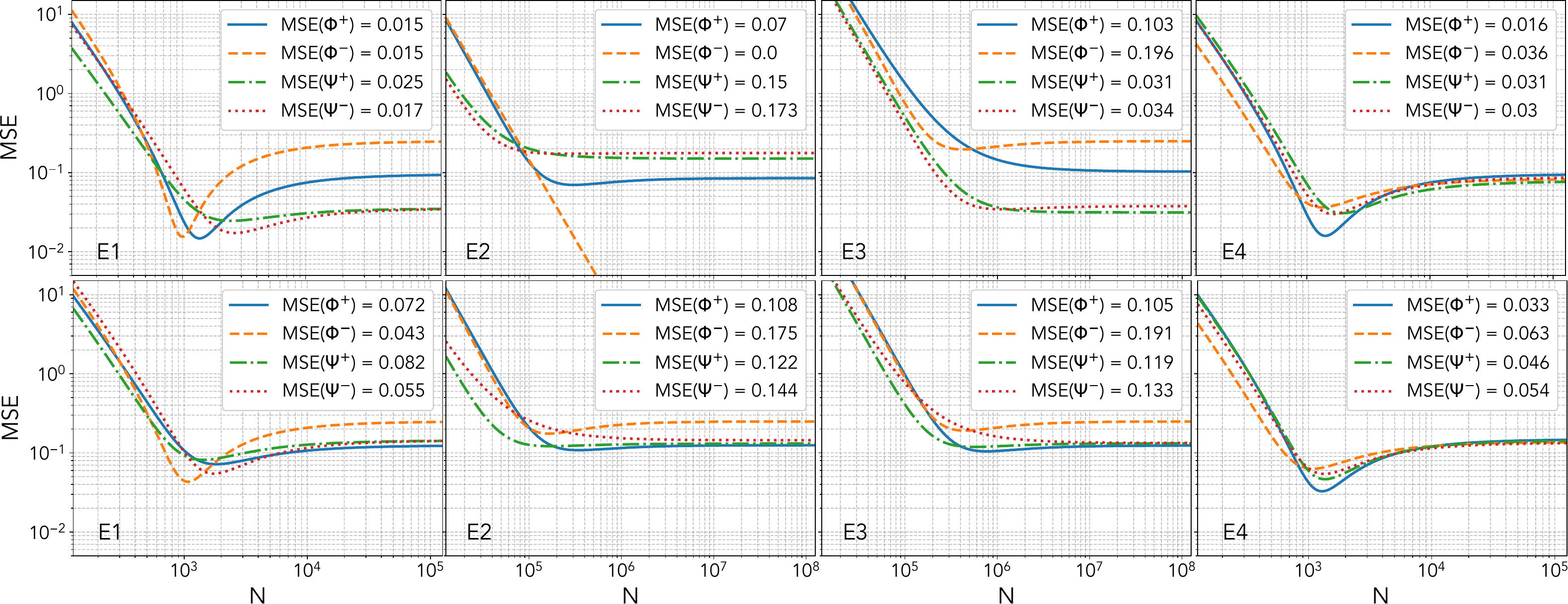}
    \caption{
    \textbf{Estimation errors with shadow tomography VS the input statistics.}
    MSEs for the configurations \sfE1, \sfE2, \sfE3, \sfE4, as a function of the unknown input statistics $N$, for separable (upper row) and entangled (bottom row) states, for the four Bell witnesses.
    In each case, large values of $N$ correspond to the trivial estimator $b\mapsto \hat o(b)=0$, corresponding to which the MSE approaches, for each target observable $\calO$, the average of $\Trace(\calO\rho)^2$ over the corresponding states.
    The observed linear behaviour of $\on{MSE}(\Phi^-;N)$ in the upper \sfE2 plot has the same origin: in this configuration $\langle\mathcal W_{\Phi^-}\rangle=0$ for all separable states, thus the zero estimator $b\mapsto \hat o(b)$ reproduces the correct value, and the MSE approaches it with the typical $1/N$ behaviour.}
    \label{fig:shadow_MSEs}
\end{figure}

\parTitle{Performances of theoretical model of the apparatus}
We report in~\cref{fig:shadow_MSEs} the MSEs as a function of the possible values of the unknown input statistics $N$, for the experimental configurations \sfE1, \sfE2, \sfE3, \sfE4, for the four Bell witnesses, and when applied to the subsets of separable and entangled states, separately.
Explicitly, the MSE is calculated for each target observable $\calW$ and statistics $N$, averaging over the states as
\begin{equation}
    \on{MSE}(\calW;N)=\frac{1}{N_{\rm test}}\sum_{i=1}^{N_{\rm test}}
    \left\lvert
    \sum_{b=1}^{N_{\rm out}}\hat o(b) \frac{N_b}{N}-\Trace(\calW\rho^{\rm test}_i)
    \right\rvert^2.
\end{equation}
Here, $N_{\rm test}$ is the number of test states $\rho_i^{\rm test}$ used to assess the accuracy of the shadow tomography estimators (note that with this method, there is no training dataset to use).
The estimator $\sum_b\hat o(b) \frac{N_b}{N}$ returns the best estimate for the expectation value of $\calW$, when $N_b$ counts are observed for each outcome $b$, and the overall statistics is $N$.
We use the experimental counts as values of $N_b$ for each training state $\rho_i^{\rm test}$, and try different values for the unknown value of $N$.
For large \(N\), this produces the trivial estimator $b\mapsto 0$, and thus in this limit 
\(\on{MSE}(\calW;N)\to\tfrac{1}{N_{\rm tr}}\sum_{i=1}^{N_{\rm tr}}\Trace\!\bigl(\calW\,\rho^{\rm test}_i\bigr)^2\), 
which explains the saturation observed in the MSE curves.
Alongside \(\on{MSE}(\calW;N)\) for varying \(N\), we also report 
\(\on{MSE}(\calW)\equiv \min_{N}\on{MSE}(\calW;N)\), 
which serves as a lower bound on the MSE achievable via shadow tomography.
Note that this minimum depends on the target observable, which means that it cannot be used as a reliable estimate for the true value of $N$.
The MSE estimated with the true $N$ could, therefore, be significantly higher than this lower bound.
We nonetheless choose here to report the minimum, to avoid biasing our comparison in favor of the QELM.
We find the shadow-tomography-based estimates to across the board yield MSEs significantly higher than those given via the QELM. 
For instance, in the \sfE1 configuration, for the witness \(\calW_{\Phi^+}\), we obtain via shadow tomography 
\(\on{MSE}(\calW_{\Phi^+}) \approx 0.015\) 
for separable states and 
\(\on{MSE}(\calW_{\Phi^+}) \approx 0.072\) 
for entangled states. 
In contrast, using the QELM trained solely on separable states, we achieve 
\(0.002\) for separable states and \(0.041\) for entangled states. 
When the QELM is trained on both separable and entangled states, the MSEs become \(0.009\) and \(0.017\), respectively. 
A brief summary of these numbers is given in~\cref{tab:mse_summary}.
Although these values will fluctuate across different experimental realizations, the enhanced performance of the QELM remains evident.

\begin{table}[h]
    \centering
    \begin{tabular}{c|c|c}
    \toprule
    \textbf{Method} & \textbf{MSE for separable states} & \textbf{MSE for entangled states} \\
    \hline
    Shadow Tomography & 0.015 & 0.072 \\
    QELM trained on separable & 0.002 & 0.041 \\
    QELM trained on separable\&entangled & 0.009 & 0.017 \\
    \bottomrule
    \end{tabular}
    \caption{\textbf{QELM estimation benchmark.} Comparison of MSEs obtained from shadow tomography vs the QELM approach, for the \sfE1 configuration with target observable \(\calW_{\Phi^+}\).}
    \label{tab:mse_summary}
\end{table}

\end{document}